\documentclass[11pt]{article}
\usepackage[nohead,margin=1.0in]{geometry}
\usepackage{amssymb, amsmath, amsthm}
\usepackage{color}
\usepackage{graphicx}
\usepackage{epstopdf}
\graphicspath{{./figures/}}
\usepackage{booktabs}
\usepackage{threeparttable}
\usepackage[utf8]{inputenc}
\usepackage{url}
\usepackage{booktabs}       
\usepackage{nicefrac}       
\usepackage{microtype}      
\usepackage{mathabx}
\usepackage{multirow}
\usepackage{dsfont}
\usepackage{cancel}
\usepackage{acronyms}
\usepackage{tikz}
\usepackage{float}
\usepackage{mathtools}
\usepackage{subcaption}
\mathtoolsset{showonlyrefs}

\newtheorem{theorem}{Theorem}[section]

\newtheorem{proposition}[theorem]{Proposition}

\newtheorem{remark}{Remark}[section]

\numberwithin{equation}{section}

\usetikzlibrary{spy}
\newcommand{\etal}{et al.}
\newcommand{\ie}{i.e., }
\newcommand{\eg}{e.g., }

\newcommand{\CQ}{{\mathcal{Q}}}
\newcommand{\BR}{{\mathbb{R}}}
\newcommand{\BE}{{\mathbb{E}}}

\newcommand{\CL}{{\mathcal{L}}}

\newcommand{\CD}{\mathcal{D}}

\newcommand{\CN}{\mathcal{N}}
\newcommand{\KL}{{\mathrm{KL}}}

\newcommand{\diag}{\mathrm{diag}}

\newcommand{\unsup}{{\rm u}}
\renewcommand{\sup}{{\rm s}}
\newcommand{\data}{\mathbb{B}}
\newcommand{\net}{\mathrm{F}}

\providecommand{\argmin}{\operatorname*{argmin}}

\newcommand{\xinstS}[1]{x^{ {\rm s}}_{#1}}
\newcommand{\xinstSit}[2]{x^{ {\rm s}}_{{#1},#2}}
\newcommand{\xinstU}[1]{x^{ {\rm u}}_{#1}}
\newcommand{\xinstUit}[2]{x^{ {\rm u}}_{#1,#2}}
\newcommand{\yinstS}[1]{y^{ {\rm s}}_{{#1}}}
\newcommand{\yinstU}[1]{y^{ {\rm u}}_{#1}}

\usepackage[colorlinks=true, allcolors=blue]{hyperref}

\newcommand{\Var}{\mathrm{Var}}
\newcommand{\qps}[1]{q_{\ifthenelse{\isempty{#1}}{\psi}{\psi_{#1}}}}
\newcommand{\qPs}[1]{q_{\ifthenelse{\isempty{#1}}{\Psi}{\Psi_{#1}}}}
\newcommand{\qpS}[1]{q_{\ifthenelse{\isempty{#1}}{\Psi}{\Psi_{#1}}}}

\newcommand{\boldtable}[1]{\textsf{\textbf{#1}}}

\usetikzlibrary{calc}
\newcommand*\canc[1]{%
  \mathchoice
    {\scriptstyle#1}
    {\scriptstyle#1}
    {\scriptscriptstyle#1}
    {\scriptscriptstyle#1}
}

\newcommand*\Dcancelto[2][0]{%
  \kern9pt%
  \begin{tikzpicture}[baseline=(current bounding box.center).anchor=west]
    \node[anchor=east,inner sep=2pt] (a) {#2};
    \draw[->] ($(a.north west)+(1pt,-2pt)$) -- ($(a.south east)+(0pt,2pt)$) node at ($(a.south east)+(4pt,1pt)$) {$\canc{#1}$};
\end{tikzpicture}
}

\newcommand{\blue}[1]{{#1}}

\title{Unsupervised Knowledge-Transfer for Learned Image Reconstruction\thanks{The work of R.B. is substantially supported by the i4health PhD studentship (UK EPSRC EP/S021930/1) and and from The Alan Turing Institute (UK EPSRC EP/N510129/1), and that of Z.K, S.A. and B.J. by UK EPSRC EP/T000864/1, and that of S.A. and B.J. also by UK EPSRC EP/V026259/1. AH acknowledges funding by Academy of Finland Projects 336796, 334817, 338408.}}

\author{Riccardo Barbano\thanks{Department of Computer Science, University College London, Gower Street, London WC1E 6BT, UK (\texttt{riccardo.barbano.19@ucl.ac.uk},\texttt{z.kereta@ucl.ac.uk}, \texttt{s.arridge@ucl.ac.uk})}\and {\v Z}eljko Kereta\footnotemark[2]\and Andreas Hauptmann\footnotemark[2]\,\,\footnotemark[3]\thanks{Research Unit of Mathematical Sciences; University of Oulu, Oulu, Finland (\texttt{Andreas.Hauptmann@oulu.fi})}\\ \and Simon R. Arridge\footnotemark[2] \and Bangti Jin\thanks{Department of Mathematics, The Chinese University of Hong Kong, Shatin, New Territories, Hong Kong, P.R. China (\texttt{bangti.jin@gmail.com,btjin@math.cuhk.edu.hk})}
}

\date{}

\begin{document}

\maketitle
\begin{abstract}
Deep learning-based image reconstruction approaches have demonstrated impressive empirical performance in many imaging modalities.
These approaches usually require a large amount of high-quality paired training data, which is often not available in medical imaging.
To circumvent this issue we develop a novel unsupervised knowledge-transfer paradigm for learned reconstruction within a Bayesian framework.
The proposed approach learns a reconstruction network in two phases.
The first phase trains a reconstruction network with a set of ordered pairs comprising of ground truth images of ellipses and the corresponding simulated measurement data.
The second phase fine-tunes the pretrained network to \blue{more} realistic measurement data without supervision.
By construction, the framework is capable of delivering predictive uncertainty information over the reconstructed image.
We present extensive experimental results on low-dose and sparse-view computed tomography showing that the approach
is competitive with several state-of-the-art supervised and unsupervised reconstruction techniques. Moreover, \blue{for test data distributed differently from the training data,} the proposed framework can significantly improve reconstruction quality not only visually, but also quantitatively in terms of PSNR and SSIM, \blue{when compared with learned methods trained on the synthetic dataset only}.
\end{abstract}

\section{Introduction}

In this work we develop a novel unsupervised knowledge-transfer framework for image reconstruction.
The reconstruction of an image is often formulated through a (linear) inverse problem
\[y = A x + \delta y,\]
where $y\in Y$ is a corrupted measurement, $\delta y$ is the additive noise, $x\in X$ is the image to be recovered, and the data acquisition is described by a linear forward map $A: X \to Y$, where $X$ and $Y$ are suitable finite-dimensional vector spaces.

In the past few years, \ac{DL}-based image reconstruction techniques have demonstrated remarkable empirical results, often substantially outperforming more conventional methods in terms of both image quality and computational efficiency \cite{arridge2019solving, ongie2020deep}.
In DL-based approaches, image reconstruction can be phrased as the problem of finding a \ac{DNN} $\net_\theta: Y \to X$ such that $\net_\theta(y)\approx x$, where the neural network $\net_\theta$ is parametrised by a parameter vector
$\theta$. In supervised learning the optimal parameter vector $\theta^\ast$
is learned from a set of ordered pairs $\data = \{(x_n, y_n)\}_{n=1}^N$ of ground truth images and the corresponding (noisy) measurement data by minimising a suitable loss
\begin{equation}\label{eqn:empiricalLoss}
    \CL(\theta) = \dfrac{1}{N}\sum_{n=1}^{N}\ell(\net_{\theta}(y_{n}),x_{n}),
\end{equation}
where $\ell(\net_\theta(y_n),x_n)$ measures the discrepancy between the network prediction $\net_\theta(y_n)$ and the corresponding ground truth image $x_n$, and is often taken to be the mean squared error.
Supervised learning has been established as a powerful tool to improve reconstruction quality and speed, rapidly becoming a workhorse in several imaging applications \cite{WangYe:2020}.

In order to deliver competitive performance, supervised learning may require many ordered pairs $(x_n,y_n)$, $n=1,\ldots,N$, which are unfortunately often not available in medical imaging applications since clean ground truth images are either too costly or impossible to collect.
Meanwhile, reconstruction methods learned in a scarce-data regime often fail to generalise on instances which belong to a different data distribution \cite{bickel2008learning, quinonero2009dataset}.
Moreover, even small deviations from the training data distribution can potentially lead to severe reconstruction artefacts (\ie supervised models can exhibit poor performance even for a small distributional shift). This behaviour is further exacerbated by the presence of structural changes such as rare pathologies; thereby significantly degrading the performance of supervisedly learned reconstruction methods \cite{antun2020instabilities}.
To make matters worse, such forms of deviation from the training data distribution are ubiquitous in medical imaging, owing to factors such as the change in acquisition protocols. For example, in magnetic resonance imaging (MRI), these factors include echo time, repetition time, flip angle, and inherent hardware variations in the used scanner \cite{KaraniKonukoglu:2021}; in \ac{CT}, they include the choice of view angles, acquisition time per view, and source-target separation.

Therefore, there is an imperative need to develop learned image reconstruction techniques that do not rely on a large amount of high-quality ordered pairs of training data. In a recent review \cite{WangYe:2020}, this issue has been identified as one of the key challenges in the next generation of learned reconstruction techniques. To address this outstanding challenge, in this work we develop a novel \ac{UKT} strategy to transfer acquired ``reconstructive knowledge'' across different datasets using the Bayesian framework.
It comprises of two phases. The first phase is supervised and is tasked with pretraining a \ac{DNN} reconstructor on data pairs of ground truth images and corresponding measurement data (which can be either simulated or experimentally collected).
The goal of this step is to capture inductive biases of the given reconstruction task using simulated or experimental data.
The second phase is unsupervised. It fine-tunes the reconstructor learned in the first phase on clinically-realistic measurement data, using a novel regularised Bayesian loss.
This fine-tunes the network to the target reconstruction task while maintaining the prior knowledge learned in the first step.
Note that unlike supervised or semi-supervised learning, the proposed framework does not assume any ground truth data from the target domain, and hence it is an unsupervised learning method.
Extensive numerical experiments with low-dose and sparse-view \ac{CT} on two datasets, i.e., \blue{FoamFanB} \cite{Pelt:2018} and LoDoFanB \cite{leuschner2019lodopab},
indicate that the proposed approach is competitive with state-of-the-art methods both quantitatively and qualitatively, and that test-time adaptation can significantly boost performance.

In summary, the development of an unsupervised knowledge-transfer framework for learned image reconstruction, and its validation on clinically realistic simulated measurement data, represent the main contributions of this work.
To the best of our knowledge, this is the first work to propose Bayesian unsupervised knowledge-transfer for test-time adaptation of a learned image reconstruction method.
Furthermore, the use of the Bayesian framework allows capturing predictive uncertainty of the obtained reconstructions. 
Our framework has the following distinct features: (i) adapting to unseen measurement data without the need for ground truth images; (ii) leveraging reconstructive properties learned in the supervised phase for effective feature representation; (iii) providing uncertainty estimates on the reconstructed images. These features make the framework very attractive for performing learned reconstruction without ordered pairs from the target domain, as confirmed by the extensive numerical experiments in Section \ref{sec:result}. The Bayesian nature of the framework is noteworthy in the emerging field of scalable uncertainty quantification for image reconstruction, where the heavy computational cost is often deemed as one of the major hurdles \cite{BarbanoArridgeJinTanno:2021}.
In contrast, the approach presented in this work is highly scalable, by building upon recent advances in variational inference \cite{gal2016uncertainty}, and hence holds significant potential for medical image reconstruction.

\subsection{Related Work}

The lack of (a sufficient amount of) reference training data has only recently motivated the development of deep learning-based image reconstruction approaches that do not require ground truth images. We identify two main groups of current learned approaches: test-time adaptation, and unsupervised approaches. 

Test-time adaptation focuses on learning under differing training and testing distributions. It often consists of fine-tuning a pretrained \ac{DNN} for a single datum at a time, or for a small set of test instances. In \cite{HanYooYe:2018, DarCukur:2020} this paradigm is used for MRI reconstruction, where reconstructive properties acquired by a network that has been pretrained on a task for which a large dataset is available, are transferred to a different task where the supervised data is scarce (but still available).
The proposed approach extends the aforementioned work from the supervised target reconstruction task to an unsupervised one.
In the context of object recognition, Sun \etal{} \cite{SunWangHardt:2020} propose to adapt only a part of a \ac{CNN} according to a self-supervised loss defined on the given test image to address distributional shift.
The model is then trained via multi-task learning, where shared features are learned jointly over supervised and self-supervised data.
Gilton \etal{} \cite{gilton2021model} adapt a pretrained image reconstruction network to reconstruct images from a perturbed forward model using only a small collection of measurements, by enforcing the data fidelity while penalising the deviation of the network parameters from the parameters of the pretrained model.
Conceptually speaking, our study is complementary to these studies. The proposed approach can be interpreted as conducting unsupervised test-time adaptation for distributional shift of the image data, but within a Bayesian framework.
Furthermore, the use of the Bayesian framework brings several distinct advantages: (i) it allows deriving the training loss in a principled manner; (ii) it can boost reconstructive performance; (iii) it simultaneously delivers the predictive uncertainty information associated with the reconstructions.

Meanwhile, \ac{DIP} is a representative unsupervised image reconstruction method, which achieves sample-specific performance using \acp{DNN} to describe the mapping from latent variables to high-quality images \cite{Ulyanov:2018deepimage}.
During inference the network architecture acts as a regulariser for reconstruction \cite{DittmerKluthMaass:2020, baguer2020computed}. Similarly, Zhang \etal{} \cite{ZhangWang:2020} use a U-Net model as the reconstruction network and propose to adapt the model through backpropagation by updating the parameters of a pretrained
U-Net under the guidance of data fidelity for each individual test data $y$, with no supervision, and showcase the approach on under-sampled {MRI} reconstruction.
Despite strong performance, it suffers from slow convergence (often requiring thousands of iterations), and the need for a well-timed early stopping, otherwise the network may overfit to the noise in the data.
The latter issue has motivated the use of an additional stabiliser \cite{baguer2020computed}.

Test time adaptation and DIP represent only two approaches that are most closely related to the present work. In recent years, there have been significant advances in unsupervised biomedical imaging reconstruction techniques and we refer interested readers to a recent review \cite{AkcakayaYamanYe:2021} on other approaches and references therein, which discusses many promising unsupervised methods.

The rest of the paper is structured as follows.
In Section \ref{sec:method} we describe the setting and discuss deep unrolled methods for image reconstruction and Bayesian \ac{DL}.
In Section \ref{sec:two-phase} we develop the proposed two-phase \ac{UKT} paradigm. In Section \ref{sec:result} we present experimental results for low-dose and sparse-view \ac{CT}, including several supervised and unsupervised benchmarks, and discuss the results obtained with the two-phase learning paradigm.
\blue{In Section \ref{sec:conclusion} we add some concluding remarks.}

\section{Preliminaries}\label{sec:method}

In this section we describe the fundamentals of how unrolled networks are used for image reconstruction.
We then describe the Bayesian approach for \ac{DNN}s, based on which we shall develop the proposed unsupervised knowledge-transfer strategy.

\subsection{Unrolled Networks}\label{ssec:unrolled_methods}

Unrolling is a popular paradigm for constructing a network $\net_\theta$ for image reconstruction.
The idea is to mimic well-established iterative  optimisation algorithms, e.g., (proximal) gradient descent, alternating direction method of multipliers, and primal-dual hybrid gradient method.
Namely, unrolled methods use an iterative procedure to reconstruct an image $x$ from the measurement $y$ by combining analytical model components (\eg the forward map $A$ and its adjoint $A^\top$) with data-driven components that are parameterised by the network parameters $\theta$ and learned from the training data.
The unrolled nature of the network allows seamlessly integrating
the underlying physics of the data acquisition process into the design of the network $\net_\theta$, which can enable the development of high-performance reconstructors from reasonably sized training datasets \cite{Monga:2021}. More specifically, given an initial guess $x_0$ (\eg the \ac{FBP} in \ac{CT} reconstruction), we recursively compute iterates \begin{equation}\label{eqn:learnedUpdates}
x_k=\net_{\theta_k}\left(x_{k-1},\nabla \CD_{k-1}\right), \quad k=1,\ldots, K,
\end{equation}
with
$$ \nabla \mathcal{D}_{k-1}:=\nabla \dfrac12\|Ax_{k-1}-y\|^2= A^\top\left(Ax_{k-1}-y\right),$$
being the gradient of the data fidelity term, where $K\geq1$ is the total number of iterations, $\net_{\theta_k}$ is the sub-network used at the $k$-th iteration, and $\theta_k$ is the corresponding weight vector.
The overall iterative process can then be written as
\[
x_K = \net_\theta\left(x_0, \nabla \CD_{0}\right),
\]
where $x_K$ is the final reconstruction, and $\net_\theta$ is the overall network, with parameters $\theta=(\theta_1,\dots,{\theta_K})$, constructed as a concatenation of sub-networks $\net_{\theta_1},\dots,  \net_{\theta_K}$. In practice, the parameters $\theta_k$ of each sub-network $\net_{\theta_k}$ can be shared across different blocks (\ie $\theta_1=\cdots=\theta_K$), \blue{a procedure known as weight-tying or weight-sharing. This allows to reduce} the total number of trainable parameters, so as to facilitate the training process. \blue{By slightly abusing the notation, we denote the shared parameter by $\theta$. In this work, we only consider the case of weights shared across the blocks, but the proposed framework extends straightforwardly to the general case.}

\subsection{Bayesian Neural Networks}\label{ssec:bayesian}

We briefly describe \acp{BNN}, in which network parameters $\theta$ are treated as random variables and are learned through a Bayesian framework so as to facilitate uncertainty quantification of the network prediction. Bayesian learning provides a principled yet flexible framework for knowledge integration, and allows quantifying predictive uncertainties associated
with a particular point estimate \cite{gal2016uncertainty,BarbanoArridgeJinTanno:2021}. Bayesian learning is ideally suited for deriving a proper training loss for combining the knowledge across different ``domains'', to which the framework proposed in Section \ref{sec:two-phase} belongs. Nonetheless, the use of \acp{BNN} for medical imaging is still not widespread due to the associated computational challenge.

In a \ac{BNN}, by placing a prior distribution $p(\theta)$ over the network parameters $\theta$ (which is commonly taken to be the standard Gaussian distribution), and by combining it with a likelihood function $p(\data|\theta)$ of the data $\data$ using Bayes' formula, we obtain a posterior distribution $p(\theta|\data)$ over the parameters $\theta$, given the data $\data$
\begin{equation*}
    p(\theta|\data) = Z^{-1} p(\data|\theta)p(\theta),
\end{equation*}
where $Z=\int p(\data|\theta)p(\theta){\rm d}\theta$ is the normalising constant. The likelihood $p(\data|\theta)$ is fully specified upon properly modelling the data noise statistics and the data generation process (\eg forward operator $A$). The posterior distribution $p(\theta|\data)$ represents the complete Bayesian solution of the learning task.

The posterior $p(\theta|\data)$ is often computationally intractable, since the computation of the normalising constant $Z$ involves a  high-dimensional integral.
To circumvent this computational issue, we adopt \ac{VI} \cite{jordan1999introduction}, which
employs the \ac{KL} divergence to construct an approximating distribution $q(\theta)$. Recall that the $\KL$ divergence $\KL[q(\theta)\|p(\theta)]$ between $q$ and $p$ is defined by
\[\KL\left[q(\theta)\|p(\theta)\right] = \int q(\theta)\log \dfrac{q(\theta)}{p(\theta)}{\rm d}\theta.\]
\ac{VI} looks for an easy-to-compute approximate posterior distribution $q_{\psi}$ parametrised by variational parameters $\psi$.
The approximation $q_{\psi}(\theta)$ is most commonly taken from a variational family consisting of products of independent Gaussians
\begin{align*}\label{eqn:VI_fam}
\CQ := \left \{q_\psi(\theta) = \prod_{j=1}^D \CN(\theta_{j}; \mu_j,\sigma^2_j) \Big| \psi\in\big(\mathbb{R}\times\mathbb{R}_{\geq0}\big)^D\right \},
\end{align*}
where the notation $\CN(\theta_{j}; \mu_j,\sigma^2_j)$ denotes a Gaussian distribution with mean $\mu_j$ and variance $\sigma_j^2$, $\psi=((\mu_j,\sigma^2_j))_{j=1}^D$ are the variational parameters, and $D$ is the total number of parameters in $\net_{\theta}$. In the literature this is commonly known as the mean field approximation.
\ac{VI} constructs an approximation $q_{\psi^\ast}(\theta)$ within the family $\CQ$ by
\begin{equation}\label{eqn:KLminimisation}
q_{\psi^\ast}(\theta)\in \argmin_{q_{\psi}(\theta)\in \CQ}\KL\left[q_{\psi}(\theta)\|p(\theta|\data)\right].
\end{equation}
Given a learned approximate posterior $q_{\psi^\ast}(\theta)$, the predictive distribution $q_{\psi^\ast}(x|y_q)$ of the target image $x$ for a new query measurement $y_{q}$ is given by $$q_{\psi^\ast}(x|y_{q}) = \int p(x|y_{q}, \theta)q_{\psi^\ast}(\theta) {\rm d}\theta.$$
A point estimate of the image $x$ can then be obtained via Monte Carlo (MC) sampling as
\[
\mathbb{E}[x]= \int x q_{\psi^\ast}(x|y_{q}){\rm d} x \approx \dfrac{1}{T} \sum_{t=1}^T \net_{\theta^t}(x_{q, 0}, \nabla\CD_{q, 0}),
\]
with $T$ Monte Carlo samples $\theta^t$, with $t=1,\ldots, T$, distributed according to $q_{\psi^\ast}(\theta)$.

When the network densities are shared across the iterates, we have
$$\net_{\theta\sim q_\psi^{\otimes K}(\theta)} :=
\net_{\theta_K \sim q_{\psi}(\theta)}\circ\cdots\circ \net_{\theta_1 \sim q_{\psi}(\theta)},$$
with the superscript $\otimes K$ denoting the $K$-fold product, and the overall iterative process reads $$ x_{K} = \net_{\theta
\sim q_\psi^{\otimes K}(\theta)}\left(x_{0}, \nabla \CD_{0}\right).$$

Note that the standard mean field approximation doubles the number of trainable parameters, which brings significant computational challenges. In practice, the training of fully Bayesian models is often non-trivial, and the performance of the resulting network is often inferior to non-Bayesian networks \cite{Osawa:2019}. \acp{BNN} are thus still not widely used in learned image reconstruction \cite{BarbanoArridgeJinTanno:2021}.
To make our approach competitive with non-Bayesian methods, while retaining the benefits of Bayesian modelling, we can adopt the strategy of \textit{being Bayesian only a little bit} \cite{Barbano:2020,Daxberger2021}. That is, we use \ac{VI} only on a subset of the parameters $\theta$, and use point estimates for the remaining parameters (or equivalently, a Dirac distribution). This can reduce the number of trainable parameters, and hence greatly facilitate the training process, while maintaining the Bayesian nature of the learning algorithm.

\begin{remark}
Apart from \ac{VI} there are other approximate inference schemes, such as MC dropout
\cite{GalGhahramani:2016} and Laplace approximation \cite{Mackay1992Thesis,Daxberger2021}. MC dropout has been widely used for modelling uncertainty, and has also found application in the medical imaging community (\eg segmentation \cite{tanno2017bayesian}), due to its computational efficiency and easy implementation, but its approximation accuracy tends to be inferior to VI. For example, MC dropout tends to severely underestimate predictive uncertainty \cite{laves2021recalibration}. Laplace approximation \cite{Mackay1992Thesis} has started to attract renewed interest, but has not been explored within medical image reconstruction so far, since the computational cost of approximating the Hessian of the loss with respect to the network parameters $\theta$ is often prohibitively high in the context of image reconstruction and a scalable and yet accurate approximation of the Hessian is still under development.
\end{remark}

\begin{remark}
\blue{In light of the decoder-encoder structure of the U-Net that is used below (cf. Fig. \ref{fig:diagram}(b) for a schematic illustration), the idea of ``being Bayesian a little bit" resembles a hybridisation of an autoencoder \cite{Vincent:2010} and a variational autoencoder \cite{KingmaWelling:2013}, which are used for the decoder and encoder parts, respectively. However, there is a major difference between the two approaches: the formulation we employ for image reconstruction is conditional on the measurement, whereas the standard autoencoder and variational autoencoder formulations are unsupervised in that they access only samples of images.}
\end{remark}

\section{Two-Phase Learning}\label{sec:two-phase}

\begin{figure}[hbt!]
  \begin{subfigure}[t]{\linewidth}
        \centering
        \includegraphics[width=0.7\textwidth]{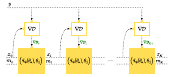}
        \caption{Diagram of the reconstructive pipeline. Each sub-network $\net_{\theta_k}$ receives as input $x_{k-1}$, $\nabla \CD_{k-1}$, $m_{k-1}$, and outputs $x_{k}$ and $m_{k}$. The gradient of the data fidelity term $\nabla \CD_{k}$ (colour-coded in green) is not an output of the sub-network, and is instead computed using the refined estimate $x_{k}$ (and the forward and adjoint operators), and then passed as input to the subsequent sub-network together with $x_{k}$ and $m_{k}$.}\label{subfig-overall}
    \end{subfigure}
    \par\bigskip
    \begin{subfigure}[t]{\linewidth}
        \centering
        \includegraphics[width=0.75\textwidth]{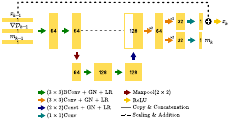}
        \caption{The architecture of $\net_{\theta_{k}}$ is a downscaled version of a residual U-Net with two scales of 64 and 128 channels.
        Each box corresponds to a multi-channel feature map, with the number of channels indicated inside. The inputs $x_{k-1}$, $\nabla \CD_{k-1}$, and $m_{k-1}$ go through a contractive path of repeated applications of two Bayesian convolutional layers (BConv), each followed by group normalisation (GN) \cite{wu2018group} and leaky ReLU (LR), with a maxpool operation in between. Maxpool halves the feature channels resulting in a coarser scale. The expansive path consists of a transposed convolution (Convt) with stride length 2, which doubles the number of feature channels. The resulting feature map is then concatenated with the feature map from the contracting path, which is further processed through a convolutional pipeline.
        The architecture then bifurcates into two identical convolutional pipelines with feature maps reduced to a single channel.
        The output of the first pipeline is added as a residual update to the initial input iterate, and projected onto the positive set to produce a new iterate $x_{k}$. The second output is the intermediate estimate of the variance. At the final iteration, we have $x_K\rightarrow \net^{\mu}_{\theta}$, and $m_K \rightarrow \net^{\sigma}_{\theta}$.
        The arrows denote different operations, and the ones which have a symbol ``$\times\!2$'' next to the arrow imply that the operation in question is repeated twice.}\label{subfig-architecture}
    \end{subfigure}
\caption{ (a) Schematic illustration of the overall iterative reconstructive process, and (b) the architecture of each sub-network.}
\label{fig:diagram}
\end{figure}

In this section we describe our novel two-phase \ac{UKT} strategy aimed at addressing the challenges associated with the lack of sufficient supervised training data in the target reconstruction task.
We systematically develop the learning strategy within a Bayesian framework with a sub-network $\net_{\theta_k}$ being a downscaled version of a residual U-Net \cite{ronneberger2015u} (cf.\ Fig.~\ref{fig:diagram}(b) for a schematic illustration), which is a popular choice in learned image reconstruction \cite{jin2017deep}, and will be used in the experiments in Section \ref{sec:result}.
The network adopts a multi-scale encoder-decoder structure consisting of an encoding component and a decoding component, whose parameters are denoted respectively
by $\theta_{{\rm e}}\in\mathbb{R}^{ D_{\rm e}}$ and $\theta_{{\rm d}}\in
\mathbb{R}^{ D_{\rm d}}$, and $\theta=(\theta_{\rm e},\theta_{\rm d})$.
In the derivation of the proposed UKT framework below, we use \ac{VI} only on the network parameters $\theta_{{\rm e}}$ of the encoder component, which can be interpreted as choosing an approximate posterior $q_{\psi^\ast}(\theta_{\rm e})$ for the encoder $p(\theta_{{\rm e}}|\data)\approx q_{\psi^{\ast}}(\theta_{\rm e})$.
The decoder parameters $\theta_{{\rm d}}$ remain deterministic, and are treated as point-estimates. The adaptation of the \ac{UKT} framework to other network architectures is straightforward.

Now we briefly describe the two phases of the proposed learning strategy. The first phase is supervised, and employs a given training dataset $\data^{\sup}=\{(\xinstS{n},\yinstS{n})\}_{n=1}^{N^{\sup}}$ where each pair $(\xinstS{n},\yinstS{n})$ consists of a ground truth image $\xinstS{n}$ and the corresponding (noisy) measurement datum $\yinstS{n}$, which can be either simulated or experimentally collected (if available).
The goal of this phase is to pretrain a reconstruction network $\net_\theta$ by learning the (approximate) posterior distribution $q_{\psi^\ast}(\theta_{\rm e})$ for the parameters $\theta_{\rm e}$ of the encoder, and the optimal deterministic parameters $\theta_{\rm d}^\ast$ of the decoder, in order to assist the unsupervised phase.
Specifically, we aim to achieve two objectives: (i) identify a sensible region for the network parameters; (ii) learn robust representations that are not prone to overfitting. \blue{Ideally, to facilitate the reconstruction quality this phase should mimic the setting of the target reconstruction task as close as possible in terms of the geometry of image acquisition (e.g., size of images and distribution of image features), and the noise statistics (e.g., distribution and noise level).}
This phase would allow learning adequate inductive biases and task-specific priors so as to enable successful subsequent unsupervised learned image reconstruction.

The second phase is unsupervised, and has access to a dataset $\data^{\unsup}=\{y_n^\unsup\}_{n=1}^{N^{\unsup}}$ which consists of only a few measurements (\eg clinically-realistic \ac{CT} sinograms), but with no access to corresponding ground truth images.
Moreover, the distribution of the measurement data in $\data^\unsup$ may differ significantly from that in $\data^\sup$.
The aim of this phase is to fine-tune the parameters $\theta$ of the reconstruction network $\net_\theta$ so that it performs well on the data $\data^\unsup$ from the target domain. This is achieved by initialising the
parameters $(\psi,\theta_{\rm d})$ of the reconstruction network $\net_\theta$ to the optimal configuration $(\psi^\ast, \theta^\ast_{{\rm d}})$ found in the first phase, and then minimising a novel loss function, which we shall derive below in the Bayesian framework. Through this phase we address the need for adaptivity to the target reconstruction task due to a potential distributional shift of the data and effectively use the inductive bias to assist the reconstruction of the target task.

\subsection{Pretraining via Supervised Learning}\label{sec:pt_sl}

In this first phase, we have access to a training dataset $\data^\sup=\{(\xinstS{n},\yinstS{n})\}_{n=1}^{N^\sup}$ of ordered pairs (which can be either simulated or experimentally collected), and we employ the Bayesian framework described in Section \ref{ssec:bayesian} to find the optimal distribution $q_{\psi^\ast}^{\sup}(\theta_{\rm e})$ for the parameters $\theta_{\rm e}$ of the encoder, which approximates the true posterior $p(\theta_{\rm e}|\data^\sup)$ and the optimal decoder parameters $\theta_{\rm d}^\ast$.
To construct the posterior $p(\theta_{\rm e}|\data^\sup)$, we first set the prior $p(\theta_{\rm e})$ over the encoder parameters $\theta_{\rm e}$ to the standard Gaussian $\CN(\theta_{\rm e}; 0, I)$, which is a standard practice in the Bayesian \ac{DL} community.
Following the heteroscedastic noise model \cite{nix1994estimating}, the likelihood $p(\xinstS{n}|\yinstS{n},\theta)$ is set to
\begin{equation}\label{eqn:likelihood}
p(\xinstS{n}|\yinstS{n},\theta)= \CN\big(\yinstS{n}; \net_{\theta}^\mu(\xinstSit{n}{0}), \hat\Sigma_n\big),
\end{equation}
with $\net_{\theta}^\mu(\xinstSit{n}{0})=\net_{\theta}^\mu\left(x_{n,0},\nabla \CD_{n, 0}, m_{n, 0}\right)$ and $\hat\Sigma_n =\diag\left(\net_{\theta}^\sigma(\xinstSit{n}{0})\right)$. Analogously, note that  $\net_{\theta}^\sigma(\xinstSit{n}{0}) = \net_{\theta}^\sigma\left(x_{n,0},\nabla \CD_{n, 0}, m_{n,0}\right)$.
Note that the network $\net_\theta$ has two outputs: the mean $\net_{\theta}^\mu$, and the variance $\net_{\theta}^\sigma$.
Here $\xinstSit{n}{0}$ denotes the initial guess used by the learned reconstruction method for the $n$-th training pair $(\xinstS{n},\yinstS{n})$. For example, in \ac{CT} reconstruction, we customarily take $\xinstSit{n}{0}$ to be the \ac{FBP}.  \blue{We refer readers to Fig.~\ref{fig:diagram}(a) for a schematic illustration, where $x_k$ and $m_k$ denote the mean and variance estimates at the $k$-th iteration, respectively.}
Up to an additive constant independent of the arguments, we can write
\[
\log p(\xinstS{n}|\yinstS{n},\theta)=-\dfrac{1}{2} \|\hat\Sigma_n^{-\frac12} (\xinstS{n}-\net_{\theta}^\mu(\xinstSit{n}{0}))\|^2-\dfrac{1}{2}\log(\det(\hat\Sigma_n)).
\]

The minimisation of \ac{KL} divergence in \eqref{eqn:KLminimisation} can be recast as the minimisation of the following loss over the admissible set $\mathbb{R}^{D_{\rm d }}\times\CQ$
\begin{equation*}
    \CL^\sup(\theta_{\rm d}, q_\psi(\theta_{\rm e})) = -\dfrac{1}{N^\sup}\sum_{n = 1}^{N^\sup} \BE_{q_{\psi}(\theta_{\rm e})}\left[ \log p\left( \xinstS{n}| \yinstS{n}, \theta \right)  \right] + \beta \KL \left [ q_{\psi}(\theta_{\rm e}) \| p(\theta_{\rm e}) \right ],
\end{equation*}
where $\beta>0$ is a regularisation parameter. This loss coincides with the negative value
of the \ac{ELBO} in \ac{VI} (when $\beta=1$).
Upon expanding the terms,
fixing the prior at $p(\theta_{\rm e})=\mathcal N(\theta_{\rm e};0,I)$, and ignoring additive constants independent of $\theta_{\rm d}$ and $q_\psi(\theta_{\rm e})$, we can rewrite the loss as \blue{(recall that $D_{\rm e}$ denotes the dimensionality of the encoder parameter $\theta_{\rm e}$)}
\begin{align}
  \CL^\sup(\theta_{\rm d},q_\psi (\theta_{\rm e})) &=\dfrac{1}{N^\sup}\sum_{n = 1}^{N^\sup} \BE_{q_{\psi} (\theta_{\rm e})}\left[\dfrac{1}{2}\|\hat\Sigma_n^{-\frac12} (\xinstS{n}-\net_{\theta}^\mu(\xinstSit{n}{0}))\|^2+\dfrac{1}{2}\log(\det(\hat\Sigma_n))\right] \nonumber\\
  &\quad + \beta \sum_{j=1}^{D_{\rm e}}\big[-\log \sigma_j + \dfrac{1}{2}(\sigma_j^2 + \mu_j^2)\big],
  \label{eqn:supervised_loss}
\end{align}
where the vector $\psi = ((\mu_j,\sigma_j^2))_{j=1}^{D_{\rm e}}$ refers to variational parameters of the approximate distribution $q_\psi(\theta_{\rm e})$, where $\mu_j$ and $\sigma_j^2$ are respectively the mean and the variance of the $j$-th component of the encoder parameters $\theta_{\rm e}$.
Note that the term $\KL \left [ q_{\psi}(\theta_{\rm e})
\| p(\theta_{\rm e}) \right ]$ affects only the encoder parameters $\theta_{e}$, whereas the decoder parameters $\theta_{\rm d}$ are treated deterministically (without any explicit penalty). In order to minimise the loss $ \CL^\sup$ with respect to the variational parameters $\psi$, we need to compute the gradient
$\nabla_{\psi} \CL^\sup$ of the loss $\CL^\sup$ with respect to $\psi$. This can be  done efficiently using the local reparametrisation trick
\cite{kingma2015variational}, which employs a deterministic dependence of the \ac{ELBO} with respect to $\psi$.

The combination of the unrolled network with Bayesian neural networks allows quantifying the uncertainty over the reconstructed image by unrolling methods, and we have termed the resulting approach (when trained in a greedy manner) as \ac{BDGD} in prior works \cite{Barbano:2020,barbano2020quantifying}. \ac{BDGD} provides natural means to quantify not only the predictive uncertainty associated
with a given reconstruction, but also to disentangle the sources from which the predictive uncertainty arises. Uncertainty is typically categorised into aleatoric and epistemic uncertainties \cite{Kendall:2017, BarbanoArridgeJinTanno:2021}.
Epistemic uncertainty arises from the uncertainty over the network parameters, and is captured by the posterior $q_{\psi}(\theta_{\rm e})$
\cite{blundell2015weight, Kendall:2017}. Aleatoric uncertainty is instead caused by the randomness in the data acquisition process. To account for this, in the loss \eqref{eqn:likelihood} we employ a heteroscedastic noise model
\cite{nix1994estimating}, which sets the likelihood $p(\xinstS{n}|\yinstS{n},\theta)$ to be a Gaussian distribution, with both its mean $\net_\theta^\mu$ and variance $\net_\theta^\sigma$ predicted by the network $\net_{\theta}$.
Accordingly, we adjust the network architecture by bifurcating the decoder output. Namely, sub-network outputs $\net_{\theta_{k}}^\mu$ are used to update the estimate $x_{k}$, whilst the intermediate term $m_{k}$, which embodies a form of ``information transmission'', is given by $\net_{\theta_{k}}^\sigma$. At the final iteration $m_K$ provides an estimate of the variance component of the likelihood; see Fig. \ref{fig:diagram}(a) again for a schematic illustration on the overall workflow of the network $\net_\theta$.

Following \cite{depeweg2018decomposition}, we can decompose the (entry-wise) predictive variance $\Var[x]$ into a sum of aleatoric ($\Delta_{{\rm A}}[y_{q}]$) and epistemic ($\Delta_{{\rm E}}[y_{q}]$)
uncertainties using the law of total variance as follows
\begin{equation*}\label{eqn:uncertainty_decomposition}
    \Var[x] = \mathbb{E}_{q_{\psi^\ast}(\theta_{\rm e})}[\Var(x|y_{q},\theta)] + \Var_{q_{\psi^\ast}(\theta_{\rm e})}[\mathbb{E}(x|y_{q}, \theta)]=: \Delta_{{\rm A}}[y_{q}] + \Delta_{{\rm E}}[y_{q}].
\end{equation*}
Upon denoting the initial guesses for the mean and the variance for a query data $y_q$ by $x_{q,0}$ and $m_{q,0}$, respectively, and abbreviating $\net^\sigma_{\theta^t}(x_{q, 0},\nabla\CD_{q, 0}, m_{q, 0})$ as $\net^\sigma_{\theta^t}(x_{q, 0})$, and $\net_{\theta^t}^\mu(x_{q, 0},\nabla\CD_{q, 0}, m_{q, 0})$ as $\net_{\theta^t}^\mu(x_{q, 0})$, we estimate $\Delta_{{\rm A}}[y_{q}]$ and $\Delta_{{\rm E}} [y_{q}]$ by $T\geq1$ Monte Carlo samples  $\{\theta_{\rm e}^t\}_{t=1}^{T}\sim q_{\psi^\ast}^{\otimes K}(\theta_{\rm e})$ as
\begin{align*}\label{eqn:uncertainty_decomposition_2}
    \Delta_{{\rm A}}[y_{q}]\approx \dfrac{1}{T} \sum_{t=1}^{T} \net^\sigma_{\theta^t}(x_{q, 0})\quad \mbox{and}\quad
    \Delta_{{\rm E}}[y_{q}]\approx\dfrac{1}{T}\sum_{t=1}^{T}\net_{\theta^t}^\mu(x_{q, 0})^{2}-\Big(\dfrac{1}{T}\sum_{t=1}^{T}\net_{\theta^t}^\mu(x_{q, 0})\Big)^{2},
\end{align*}
where all the operations on vectors are understood entry-wise.

\begin{remark}
There are at least two alternative loss functions that can be derived from the Bayesian loss \eqref{eqn:supervised_loss}.
The first option is to set the parameters $\theta$ as fully deterministic, which gives rise to the following non-Bayesian loss
\begin{equation*}
  \CL^\sup(\theta) = \dfrac{1}{N^\sup}\sum_{n = 1}^{N^\sup} \big[\dfrac{1}{2}\|\hat\Sigma_n^{-\frac{1}{2}} (\xinstS{n}-\net^\mu_{\theta}(\xinstSit{n}{0}))\|^2+ \dfrac{1}{2}\log(\det(\hat\Sigma_n))\big]+ \dfrac{\beta}{2}\|\theta_{\rm e}\|^2.
\end{equation*}
Note that this loss does not penalise the decoder parameters $\theta_{\rm d}$, as in the Bayesian formulation, whereas it penalises the encoder parameters $\theta_{\rm e}$ by the standard weight decay, which corresponds directly to the
standard Gaussian prior $p(\theta_{\rm e})=\mathcal{N}(\theta_{\rm e};0,I)$ on the encoder parameters $\theta_{\rm e}$.
The presence of the log-determinant $\log(\det(\hat\Sigma_n))$ is due to heteroscedastic noise modelling \cite{nix1994estimating}, and accordingly the network $\net_\theta$ has two outputs, one for the mean and the other for the variance.
The second option is to fix the output noise variance as $\hat\Sigma_n=\sigma^2 I$ $($with known $\sigma)$ in the heteroscedastic noise modelling. This leads to the following loss
\begin{equation*}
  \CL^\sup(\theta) =\dfrac{1}{N^\sup}\sum_{n = 1}^{N^\sup} \dfrac{1}{2\sigma^2}\|\xinstS{n}-\net^\mu_{\theta}(\xinstSit{n}{0})\|^2+ \dfrac{\beta}{2}\|\theta_{\rm e}\|^2.
\end{equation*}
This is essentially identical to the loss in \eqref{eqn:empiricalLoss} (modulo weight decay), which is arguably the most popular loss for obtaining supervised end-to-end DL-based image reconstruction algorithms.
\end{remark}

\subsection{Unsupervised Knowledge-Transfer}\label{sect:ukt}

In the second phase we use the Bayesian framework to integrate the knowledge learned in the first phase to new imaging data for which we don't have access to paired training data but only to noisy observations.
Note that the knowledge of the trained network (on the supervised data $\data^\sup$) is encoded indirectly in the posterior distribution $q^\sup_{\psi^\ast}(\theta_{\rm e})$ and in the optimal parameters $\theta_{\rm d}^\ast$.
The goal of the second phase is to approximate the true posterior $p(\theta_{\rm e}|\data^\sup, \data^\unsup)$, and to find the updated optimal decoder parameters $\theta^\ast_{\rm d}$ given the measurement data $\data^\unsup$ and the supervised data $\data^\sup$ from the first phase.
This can be achieved as follows. By Bayes' formula, the posterior distribution $p(\theta_{\rm e}|\data^\sup, \data^\unsup)$ is given by
\[
p(\theta_{\rm e}|\data^\sup, \data^\unsup) = (Z^\unsup)^{-1}p(\data^\unsup|\theta_{\rm e})p(\theta_{\rm e}|\data^\sup).
\]
Here $p(\data^\unsup|\theta_{\rm e})$ is the likelihood at test-time (\ie  the likelihood of the measurement data $\data^\unsup$ from the target reconstruction task), and the normalising constant $Z^\unsup=\int p(\data^\unsup|\theta_{\rm e})p(\theta_{\rm e}|\data^\sup){\rm d}\theta_{\rm e}$ is the marginal likelihood of the total observed data $(\data^\sup,\data^\unsup)$.
We approximate the posterior $p(\theta_{\rm e}|\data^\sup)$ (from the supervised phase) by the estimated optimal posterior $q_{\psi^\ast}^\sup(\theta_{\rm e})$, which is learned in the first phase, thus encapsulating the ``proxy'' knowledge we have acquired from the supervised dataset $\data^\sup$.
An approximation $q_{\psi^\ast}^\unsup(\theta_{\rm e})$ to the true posterior $p(\theta_{\rm e}|\data^\sup,\data^\unsup )$ for the combined data $(\data^\sup, \data^\unsup)$ can then be obtained using \ac{VI} as
\begin{align*}
(\theta_{\rm d}, q_{\psi^\ast}^\unsup(\theta_{\rm e}))\in \argmin_{\theta_{\rm d}\in\mathbb{R}^{D_{\rm d}}, q_{\psi}(\theta_{\rm e})\in \CQ}\CL^\unsup(\theta_{\rm d},q_{\psi}(\theta_{\rm e})),
\end{align*}
where the objective function is given by
\begin{equation}\label{eqn:unsuploss}
    \CL^\unsup(\theta_{\rm d}, q_{\psi}(\theta_{\rm e})) :=\KL\left [q_{\psi}(\theta_{\rm e}) \big\| (Z^\unsup)^{-1}p(\data^\unsup|\theta_{\rm e })q_{\psi^\ast}^\sup(\theta_{\rm e})\right].
\end{equation}
The approximate posterior $q^\sup_{\psi^\ast}(\theta_{\rm e})$ over the supervised dataset $\data^\sup $ is by construction used as a prior in the second phase. It remains to construct the likelihood $p(\data^\unsup|\theta_{\rm e })$ for the unsupervised dataset $\data^\unsup$. For any measurement datum $y^\unsup\in\data^\unsup$, the likelihood $p(y^\unsup|\theta_{\rm e})$ is set to
\[
p(y^\unsup|\theta_{\rm e})= \CN(y^\unsup; A \net_{\theta}^\mu(x^\unsup_{0}), \sigma^2 I).
\]
Upon letting $\bar y^\unsup = A \net_{\theta}^\mu(x^\unsup_{0})$, we have 
\begin{equation*}
    \log p(y^\unsup|\theta_{\rm e}) = -\dfrac{1}{2 \sigma^2 }\|\bar y^\unsup - y^\unsup\|^2 -\dfrac{m}{2}\log(2\pi\sigma^2).
\end{equation*}
Note that unlike in \eqref{eqn:supervised_loss}, this likelihood would exert no influence on the component $\net^\sigma_\theta$ of the network output $\net_\theta$ (arising from the heteroscedastic modelling).
To address this, we shall, inspired by the bias variance decomposition, replace the log-likelihood $\log p(\yinstU{}|\theta_{\rm e})$ with a suitable modification.
For $p(\tilde x^\unsup)=\CN\big(x^\unsup; \net_{\theta}^\mu(x^\unsup_{0}), \hat{\Sigma}\big)$, using the standard bias-variance decomposition, we obtain
\[ \mathbb{E}_{p(\tilde x^\unsup)} [\|A\tilde x^\unsup-y^\unsup\|^2]=\|A\mathbb{E}_{p(\tilde x^\unsup)}[\tilde{x}^\unsup]-y^\unsup\|^2+\mathbb{E}_{p(\tilde x^\unsup)}[\|A\mathbb{E}_{p(\tilde x^\unsup)}[\tilde{x}^\unsup]-A\tilde{x}^\unsup\|^2].\]
By the definition of $\bar{y}^\unsup$, the first term can be rewritten as $\|\bar y^\unsup - y^\unsup\|^2$.
Meanwhile, for a random vector $w$ with mean $\mathbb{E}[w]=0$ and covariance $\mathrm{Cov}(w)$, we have  
\begin{equation*}
\mathbb{E}[\|w\|^2]=\mathbb{E} [w^\top w]={\rm trace}({\rm{Cov}}(w)).
\end{equation*}
Since $A\tilde{x}^\unsup-A\mathbb{E}_{p(\tilde x^\unsup)}[\tilde{x}^\unsup]$ is a zero mean random vector with covariance
${\rm{Cov}}(w) = A\hat{\Sigma}A^\top$, we have
\[ \mathbb{E}_{p(\tilde x^\unsup)}[\|A\mathbb{E}_{p(\tilde x^\unsup)}[\tilde{x}^\unsup]-A\tilde{x}^\unsup\|^2]={\rm trace}(A\hat{\Sigma}A^\top).
\]
Consequently,
\[
\mathbb{E}_{p(\tilde x^\unsup)} \|A\tilde{x}^\unsup-y^\unsup\|^2=\|A\hat{x}^\unsup-y^\unsup\|^2+ {\rm trace}(A\hat{\Sigma}A^\top),\quad 
\mbox{with } \hat{x}^\unsup=\net_\theta^\mu(x_0^\unsup)\]
This will be used in the loss function in the modified log-likelihood.
In practice, the term $\text{trace}(A \hat\Sigma A^\top)$ can be approximated using randomised trace estimators (\eg the Hutchinson's estimator \cite{bujanovic2021norm}).
The computation of the optimal variational parameters $\psi^\ast$ and the optimal decoder parameter $\theta_{\rm d}^\ast$ by minimising the negative value of the \ac{ELBO} proceeds analogously to the supervised phase, but with the key changes outlined above.

In addition to enforcing data fidelity,  we also include a regularisation term to the loss in \eqref{eqn:unsuploss},
\[
 \tilde{\CL}^\unsup(\theta_{\rm d}, q_\psi(\theta_{\rm e}) = \CL^\unsup(\theta_{\rm d}, q_\psi(\theta_{\rm e})) + \gamma \BE_{q_{\psi}(\theta_{\rm e})}\left[{\rm TV}(\net^{\mu}_{\theta}(x^\unsup_{0}))\right],
\]
where as a regulariser we take the total variation seminorm ${\rm TV}(u)=\|\nabla u\|_1$, and $\gamma>0$ is the regularisation parameter.
This incorporates prior knowledge over expected images by penalising unlikely or undesirable solutions.
TV is widely used in image reconstruction, due to its edge-preserving properties \cite{BrediesHoller:2020}, and has also been applied to learned reconstruction \cite{baguer2020computed,Cascarano:2020}. 
\blue{Intuitively, without the TV term, optimising the loss is akin to minimising the fidelity, and thus the training process is prone to overfitting, especially when the neural network is over-parameterised, necessitating the use of early stopping (which also has a regularising effect). The numerical experiments indicate that incorporating this term can help stabilise the training process and lead to improved reconstructions, which agrees with earlier observations \cite{baguer2020computed,Cascarano:2020}.}
In summary, the loss at the second phase reads
\begin{align}
   \tilde{\CL}^\unsup(\theta_{\rm d},q_\psi(\theta_{\rm e})) =  -\BE_{q_\psi(\theta_{\rm e})}\left[ \log p\left(\data^\unsup| \theta_{\rm e}\right) - \gamma {\rm TV}(\net^{\mu}_{\theta}(x^\unsup_{0})) \right] + \beta\KL \left [ q_{\psi}(\theta_{\rm e}) \| q^\sup_{\psi^\ast}(\theta_{\rm e}) \right ],
\end{align}
which upon expansion, relabelling, and the aforementioned modifications, leads to the loss
\begin{align}
\tilde{\CL}^\unsup(\theta_{\rm d}, q_\psi(\theta_{\rm e}))  &=\dfrac{1}{N^\unsup}\sum_{n=1}^{N^\unsup}\BE_{q_{\psi}(\theta_{\rm e})}\big[\dfrac{1}{2}\| y_n^\unsup - A\net^\mu_{\theta}(x^\unsup_{n,0}) \|^2  +  {\rm trace}(A \hat\Sigma A^\top) + \gamma{{\rm TV}(\net_{\theta}^{\mu}(x^\unsup_{n,0}))} \big]\nonumber\\
&\quad +
\beta \KL \left [ q_{\psi}(\theta_{\rm e}) \| q^\sup_{\psi^\ast}(\theta_{\rm e}) \right ].
\label{eqn:unsupervised_loss}
\end{align}
Since $q_{\psi}(\theta_{\rm e})$  and $q^\sup_{\psi^\ast}(\theta_{\rm e})$ are constructed as the products of independent Gaussians (\ie mean field approximation), the term  $\KL [ q_{\psi}(\theta_{\rm e}) \| q^\sup_{\psi^\ast}(\theta_{\rm e})]$ has a closed-form expression given by
\begin{equation*}
\KL \left[ q_{\psi}(\theta_{\rm e}) \| q^\sup_{\psi^\ast}(\theta_{\rm e}) \right] = \sum_{j=1}^{D_{\rm e}} \bigg[\log \frac{\sigma_j^\sup}{\sigma_j}+\frac{\sigma_j^2+(\mu_j-\mu_j^\sup)^2}{2(\sigma_j^\sup)^2}-\frac12\bigg],
\end{equation*}
where $\psi = \left((\mu_j,\sigma_j)\right)_{j=1}^{D_{\rm e}}$ refers to variational parameters of the approximate distribution $q_\psi(\theta_{\rm e})$, where $\mu_j$ and $\sigma_j$ are the mean and the variance of the $j$-th component of $\theta_{\rm e}$, and $\sigma^{\rm s}_j$ and $\mu^{\rm s}_j$ are the optimal variational parameters learned in the first phase (and thus fixed during the second phase). Note that the loss in \eqref{eqn:unsupervised_loss} represents only one possibility for unsupervised knowledge transfer, and there are alternatives. In the appendix, we derive an alternative training loss, by constructing the likelihood $p(y^\unsup|\theta_{\rm e})$ differently, which also allows interpreting the loss $\tilde{\CL}^\unsup$ as an approximate Bayesian loss.

It is instructive to interpret the terms in the loss $\tilde{\CL}^\unsup$ in the lens of more familiar variational regularisation \cite{EnglHankeNeubauer:1996,ItoJin:2015}.
The first term in  \eqref{eqn:supervised_loss} enforces data fidelity, which encourages the learned network $\net_{\theta}$ to be close to the right-inverse of $A$ (\ie the action of the forward map $A$ on the output of $\net_{\theta}(x^\unsup_{0})$ is close to the measurement data $y^\unsup$). The second term, ${\rm trace}(A \hat\Sigma A^\top)$, controls the growth of the variance component, and along with the first term arises naturally when performing approximate \ac{VI} (with a Gaussian likelihood) on the posterior distribution $p(\theta_{\rm e}|\data^\sup,\data^\unsup )$; see the appendix for further discussions.
Note that this term does not appear if one considers only the usual maximum a posteriori (MAP) estimator to the posterior distribution $p(\theta_{\rm e}|\data^\sup,\data^\unsup )$.
The third term, the TV regulariser, plays a crucial role in stabilising the learning process \cite{baguer2020computed}.
The fourth term $\KL[ q_{\psi}(\theta_{\rm e}) \| q^\sup_{\psi^\ast}(\theta_{\rm e})]$ forces the posterior $q_\psi(\theta_{\rm e})$ to be close to the prior $q^\sup_{\psi^\ast}(\theta_{\rm e})$ of the unsupervised phase (which is the posterior obtained during the supervised phase). These properties
together give rise to a highly flexible \ac{UKT} paradigm: the adaptation can be done individually for each query image datum (which is natural for streaming data) or for a whole batch of measurement data.
The regularisation parameters $\gamma>0$ and $\beta>0$ control the strength of the related penalty terms.
\blue{In practice, it is important to choose the regularisation parameters $\beta$ and $\gamma$ suitably, as in any inverse technique. In our experiments $\beta$ and $\gamma$ are chosen on a validation set.}

\begin{remark}
The loss $\tilde{\CL}^{\rm u}$ in \eqref{eqn:unsupervised_loss}
can be viewed as a generalisation of the more conventional non-Bayesian approaches for domain
adaptation
\begin{equation}\label{eqn:supervised_loss-map}
   \CL^{\rm u}(\theta) = \dfrac{1}{N^\unsup}\sum_{n=1}^{N^{\unsup}}\big[\dfrac{1}{2}\|\yinstU{n} - A\net_{\theta}(\xinstUit{n}{0})  \|^2 + \gamma{{\rm TV}(\net_{\theta}(\xinstUit{n}{0}))}\big] +
    \dfrac{\beta}{2}\|\theta_{\rm e}-\theta_{\rm e}^{\rm s}\|^2,
\end{equation}
where $\theta^{\rm s}_{\rm e}$ is the optimal encoder network parameter learned at the supervised phase. This loss encourages the network output $\net_\theta(\xinstU{0})$ to be close to piece-wise constant, and meanwhile, the corresponding network should not deviate too much from $\theta^{\rm s}_{\rm e}$.
Due to the use of the Bayesian framework, the \ac{UKT} loss \eqref{eqn:unsupervised_loss} involves extra terms that are related to the variance of the parameters. Formally, the loss in \eqref{eqn:supervised_loss-map} can also be obtained by considering the MAP estimator of the posterior distribution $p(\theta_{\rm e}|\data^{\rm s},\data^{\rm u} )$,
concurring with the well-known connection between the MAP estimator and the posterior distribution.
Nonetheless, even if considering the loss \eqref{eqn:supervised_loss-map} alone, the Bayesian framework elucidates the standing assumptions for obtaining the loss. The loss in \eqref{eqn:supervised_loss-map} is closely connected to the loss
\begin{equation}\label{eqn:supervised_loss-map-1}
   \CL^{\rm u}(\theta) = \dfrac{1}{N^\unsup}\sum_{n=1}^{N^{{\rm u}}}\dfrac{1}{2}\| \yinstU{n} - A\net_{\theta}(\xinstUit{n}{0})  \|^2 +
    \dfrac{\beta}{2}\|\theta-\theta^{\rm s}\|^2,
\end{equation}
which also penalises the deviation of decoder parameters $\theta_{\rm d}$ from the pretrained parameters $\theta_{\rm d}^\sup$. This is essentially the training loss employed in \cite{gilton2021model}. The other major difference between \eqref{eqn:supervised_loss-map} and \eqref{eqn:supervised_loss-map-1} lies in use of the TV penalty on the network output $\net_\theta(\xinstU{0})$.

It is also worth noting that $\hat\Sigma$ affects the loss $\tilde{\CL}^{\rm u}$ in \eqref{eqn:unsupervised_loss} only via the term ${\rm trace}(A \hat\Sigma A^\top)$, controlling the covariance of the estimate, but not via the data fidelity term. This is due to the simplified derivation of the loss. The genuine Bayesian loss in \eqref{eqn:unsupervised_loss-alt} to be derived in the appendix does incorporate $\hat\Sigma$ into noise covariance and consequently it does enter into the noise weighting matrix in the data fidelity. See the appendix for further discussions.
\end{remark}

\section{Experiments and Results}\label{sec:result}
In this section we present numerical experiments on simulated data to showcase the performance of the proposed \ac{UKT} framework.

\subsection{Experimental Settings}

{
\newcommand{\showpiclodo}[2]{%
\begin{tikzpicture}[spy using outlines={rectangle, magnification=2.25,
size=2.75cm, connect spies, red, thick}]%
\draw (0,0) node [anchor=north] {\includegraphics[width=4cm, height=4cm,  angle=-90]{../figures/#2}};%
\spy on (0.25cm, -2cm) in node at (-0.75cm, -0.5cm);
\end{tikzpicture}\hspace*{-2mm}%
}

\newcommand{\ellshowpic}[2]{%
\begin{tikzpicture}[]%
\draw (0,0) node [anchor=south, yshift=-5pt] {\phantom{f}#1\phantom{g}};%
\draw (0,0) node [anchor=north] {\includegraphics[width=4cm,  height=4cm]{../figures/#2}};%
\end{tikzpicture}\hspace*{-2mm}%
}

\newcommand{\showpicfoam}[2]{%
\begin{tikzpicture}[spy using outlines={rectangle, magnification=2.25,
size=2.75cm, connect spies, red, thick}]%
\draw (0,0) node [anchor=north] {\includegraphics[width=4cm, height=4cm,  angle=-90]{../figures/#2}};%
\spy on (0.25cm, -2cm) in node at (-0.75cm, -0.5cm);
\end{tikzpicture}\hspace*{-2mm}%
}

\begin{figure}
    \centering
    \quad
    \ellshowpic{}{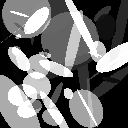}
    \quad
    \showpicfoam{}{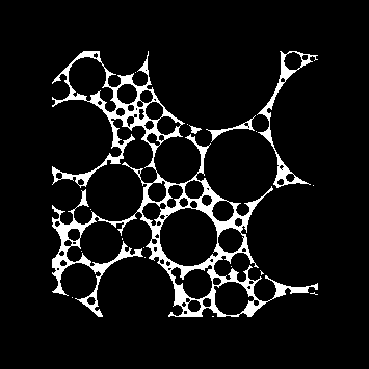}
    \quad
    \showpiclodo{}{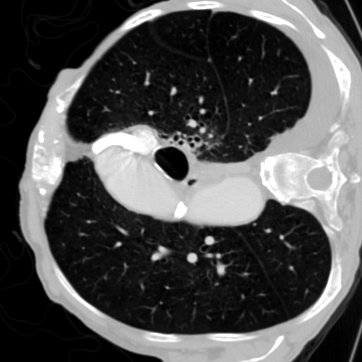}
    \quad
    \caption{Representative ground truth images from Ellipses (left), \blue{FoamFanB (middle)} and LoDoFanB (right) datasets. The window of the LoDoFanB dataset is set to a Hounsfield unit (HU) range $\approx$ [-1000, 400].}
    \label{fig:fig_data}
\end{figure}
}

First we describe the experimental setting, including datasets, data generation, benchmark methods and training details.

\smallskip
\noindent\textbf{Datasets.} In the experiments we use \blue{three} datasets: Ellipses, \blue{FoamFanB} and LoDoFanB.
The Ellipses dataset consists of random phantoms of overlapping ellipses, and is commonly
used for inverse problems in imaging \cite{adler2017solving}. The intensity of the
background is taken to be $0$, the intensity of each ellipse is taken randomly between $0.1$
and $1$, and the intensities are added up in regions where ellipses overlap.
The phantoms are of size $128\times 128$; see Fig.~\ref{fig:fig_data} for a
representative phantom. The training set contains $32\,000$ pairs of phantoms and sinograms, while the test set consists of 128 pairs. This dataset is used for the training of all the methods that involve supervised training.
\blue{
The FoamFanB dataset is constructed using a cylindrical foam phantom containing $100\,000$ randomly-placed non-overlapping bubbles. The phantom consists of $100$ slices of size $1024\times 1024$ and is generated with the open-source {\texttt{foam\_ct\_phantom}} package \cite{Pelt:2018}. 
Analytic projection images of the phantoms were also computed using the package.
Each slice is then cropped into four $256\times 256$ square sections, which are zero padded with $50$ pixels in all four directions. Out of the resulting 400 slices, we randomly retain half; see Fig.~\ref{fig:fig_data} for a representative slice. The intensity of the pixels are either $0$ or $0.5$, which allows retaining finer structures in the images.}
The LoDoFanB dataset \cite{leuschner2019lodopab} is more medically realistic, and consists of $223$ human chest \acp{CT}, in which the (original) slices from the LIDC/IDRI Database \cite{armato2004lung} have been pre-processed, and the resulting images are of size $362\times362$; see Fig.~\ref{fig:fig_data} for a representative slice.
The \blue{FoamFanB} and LoDoFanB datasets are used in the unsupervised phase, where we assume to know only the sinograms.
The ground truth images are only used to evaluate the performance of all the studied methods, unless otherwise specified.

\smallskip
\noindent\textbf{Data generation.} For the forward map $A$, taken to be the Radon transform, we employ a two-dimensional fan-beam geometry with $600$ angles for the low-dose \ac{CT} setting, and $100$ angles for the sparse-view \ac{CT} setting. Source-to-axis and axis-to-detector distances are both set to $500$ mm. For both datasets we apply a corruption process given by $\lambda \exp \left (-\mu A x\right)$,
where $\lambda \in \BR^{+}$ is the mean number of photons per pixel and is fixed at $8000$ (corresponding to low-dose \ac{CT}), and $\mu\in\BR^{+}$ is the attenuation coefficient, set to $0.2$. We linearise the forward model by applying the transformation $-\log(\cdot)/\mu$.
We can then use $\frac{1}{2}\|Ax-y\|^2$ as the data fidelity term since post-log measurements of low-dose \ac{CT} approximately follow a Gaussian distribution \cite{wang2006noise, li2004nonlinear}.

\smallskip
\noindent\textbf{Benchmark methods.} We compare the proposed \ac{BDGD}+\ac{UKT} approach with several unsupervised and supervised benchmarks. Unsupervised methods include \ac{FBP} (using a Hann filter with a low-pass cut-off 0.6), (isotropic)
TV regularisation, and \ac{DIP}+TV \cite{baguer2020computed}.
Supervised methods include U-Net based post-processing (\ac{FBP}+U-Net) \cite{chen2017low}, two learned iterative schemes: \ac{LGD} \cite{adler2017solving} and \ac{LPD} \cite{adler2018learned}, and \ac{BDGD} (\ie without \ac{UKT}) \cite{Barbano:2020,barbano2020quantifying}.
U-Net is widely used for post-processing  (\eg denoising and artefact removal), including \acp{FBP} \cite{jin2017deep}, and  our implementation follows \cite{baguer2020computed} using a slightly down-scaled version of the standard U-Net.
\ac{LGD} and \ac{LPD} are widely used, with the latter often seen as the standard benchmark for supervised deep tomographic reconstruction. \ac{BDGD} exhibits competitive performance while being a Bayesian method \cite{Barbano:2020,barbano2020quantifying}.

\begin{table*}[ht]\centering
\caption{\blue{Reconstruction methods used in this work. For each method, the number of learnable parameters is indicated, as well as approximate runtime for both low-dose \ac{CT} and sparse-view \ac{CT} on the LoDoFanB dataset, is reported.}}\label{tab:tab1}
{\renewcommand{\arraystretch}{1.25}\small \setlength{\tabcolsep}{2pt}
\resizebox{0.5\textwidth}{!}{
\begin{tabular}{cccccc}
\cline{1-2}\cline{4-4}\cline{6-6}
\multicolumn{2}{c}{\boldtable{Methods}} && Parameters && Runtime\\
\cline{1-2}\cline{4-4}\cline{6-6}
\multirow{3}{*}{{Unsupervised}} & FBP && 1 && 38ms/7ms  \\
&TV && 1 && 20s/10s\\
&DIP+TV && $2.9\cdot 10^{6}$ && 20min/18min\\
\cline{1-2}\cline{4-4}\cline{6-6}
\multirow{4}{*}{{Supervised}} & FBP+U-Net && $6.1 \cdot 10^5$ && $5$ms\\
&LGD && $1.3 \cdot 10^{5}$ && $89$ms/$34$ms\\
&LPD && $2.5 \cdot 10^{5}$ && 180ms/55ms \\
&BDGD && $8.8\cdot 10^{5}$ && 7s/6s \\
\cline{1-2}\cline{4-4}\cline{6-6}
\multicolumn{2}{c}{{BDGD+UKT}} && $8.8\cdot 10^{5}$ && 7s/6s\\
\cline{1-2}\cline{4-4}\cline{6-6}
\end{tabular}
}
}
\end{table*}

\smallskip
\noindent\textbf{Training, hyper-parameters, and implementation.} All supervised methods are first trained on the Ellipses dataset, and then tested on Ellipses, \blue{FoamFanB} and LoDoFanB datasets separately. Unless otherwise stated, the learned models are not adapted to the \blue{FoamFanB} and LoDoFanB datasets, but perform reconstruction directly on a given sinogram.
The methods were implemented in PyTorch, and trained on a GeForce GTX 1080 Titan GPU.
All operator-related components (\eg forward operator, adjoint, and \ac{FBP}) are implemented using the Operator Discretisation Library \cite{adler2017operator} with the \texttt{astra\_gpu} backend \cite{van2015astra}.

For all the unsupervised methods (\ac{FBP}, TV, \ac{DIP}+TV), the hyperparameters (frequency scaling in \ac{FBP} and regularisation parameter in TV and \ac{DIP}+TV) are selected to maximise the PSNR on a subset of the dataset consisting of 5 images.
\ac{DIP}+TV adopts a U-Net architecture proposed in \cite{baguer2020computed} (accessible in the DIVal library \cite{leuschner2019}): a 5-scale U-Net without skip connections for the Ellipses dataset, and a 6-scale U-Net with skip connections only at the last two scales for \blue{FoamFanB} and LoDoFanB datasets.
For both architectures the number of channels is set to 128 at every scale.
In Table \ref{tab:tab1} we report the number of parameters used for the LoDoFanB dataset.

All learned reconstruction methods were trained until convergence on the Ellipses dataset.
\ac{FBP}+U-Net implements a down-sized U-Net architecture with 4 scales and skip connections at each scale.
\ac{LGD} is implemented as in \cite{adler2018learned}, where the parameters of the reconstructor are not shared across the iterates, and we use $K=10$ unrolled iterations.
\ac{LPD} follows the implementation in \cite{adler2018learned}.
We train \ac{FBP}+U-Net, \ac{LGD} and \ac{LPD} by minimising the loss in \eqref{eqn:empiricalLoss} using the Adam optimiser and a learning rate schedule according to cosine annealing \cite{loshchilov2016sgdr}. All models are trained for 30 epochs. \ac{BDGD} uses a multi-scale convolutional architecture (cf.  Fig.~\ref{fig:diagram}), with $K = 3$ unrolled iterations.
Furthermore, the \ac{UKT} phase is initialised with parameters $(\psi\ast, \theta^\ast_{\rm d})$, which are obtained at the end of the supervised training on the Ellipses dataset. 
\blue{For the FoamFanB dataset, the regularisation parameter $\gamma$ is set to $5\cdot 10^{-5}$ for the low-dose setting and to $1\cdot 10^{-4}$ for the sparse-view setting.}
Analogously, for the LoDoFanB dataset, the regularisation parameter $\gamma$ is set to $1\cdot 10^{-4}$ for the low-dose setting and to $5\cdot 10^{-4}$ for the sparse-view setting.
On both datasets, $\beta$ is set to $1\cdot 10^{-4}$ for both settings. $T=10$ Monte Carlo samples are used to reconstruct the point estimate, and to compute the associated uncertainty estimates. \blue{A Pytorch implementation of the proposed approach is publicly available at \url{https://github.com/rb876/unsupervised_knowledge_transfer} to reproduce the numerical experiments.}

\subsection{Experimental Results}

In Table \ref{tab:tab2} we report PSNR and SSIM values for the studied datasets. 
We observe that unsupervised methods give higher PSNR and SSIM values on FoamFanB and LoDoFanB datasets than on the Ellipses dataset, \blue{with FBP on FoamFanB being the exception}.
The converse is true for supervised methods. Moreover, TV and \ac{DIP}+TV outperform supervised reconstruction methods in both low-dose and sparse-view \ac{CT} settings for {FoamFanB} and LoDoFanB datasets. The results for \ac{BDGD}+\ac{UKT} and \ac{BDGD} indicate that adapting the parameters on the {given} dataset allows achieving a noticeable improvement in reconstruction quality in both low-dose and sparse-view \ac{CT} settings.
Note also that \ac{BDGD}+\ac{UKT} outperforms all supervised reconstruction methods, while performing on par with \ac{DIP}+TV (but the corresponding computation time is only a small fraction of that for the latter). \blue{This last observation is not surprising, since the test data (FoamFanB and LoDoFanB) are distributed differently from the synthetic training data (Ellipses). As a result, the performance of supervised reconstruction methods deteriorates significantly.}

\begin{table*}[ht]\centering
\caption{Comparison of reconstruction methods for the Ellipses, FoamFanB, and LoDoFanB datasets by average PSNR and SSIM. All supervised methods are trained on the  Ellipses dataset. Learned models are then tested on the FoamFanB and LoDoFanB datasets. In the table, the two best performing methods are highlighted in bold case.}\label{tab:tab2}
{\renewcommand{\arraystretch}{1.25}\small \setlength{\tabcolsep}{2pt}
\resizebox{\textwidth}{!}{
\begin{tabular}{cccccccccc}
\multicolumn{2}{c}{} && \multicolumn{3}{c}{\boldtable{Low-Dose CT}} && \multicolumn{3}{c}{\boldtable{Sparse-View CT}}\\
\cline{1-2}\cline{4-6}\cline{8-10}
\multicolumn{2}{c}{\boldtable{Methods}} && Ellipses & \blue{FoamFanB} &  {LoDoFanB}  && {Ellipses} & \blue{FoamFanB} & {LoDoFanB} \\
\cline{1-2}\cline{4-6}\cline{8-10}
\multirow{3}{*}{{Unsupervised}} & FBP && 28.50/0.844 & \blue{20.73/0.629} & 33.01/0.842 && 26.74/0.718 & \blue{16.34/0.174} & 29.10/0.593 \\
&TV && 33.41/0.878 & \blue{36.39/0.939}  & 36.55/0.869 && 30.98/0.869 & \blue{27.53/0.832} & 34.74/0.833 \\
&DIP+TV && 34.53/0.957 & \blue{\textbf{38.42}/\textbf{0.997}} & \textbf{39.32}/\textbf{0.896} && 32.02/0.931 & \blue{\textbf{31.99}/\textbf{0.987}} & \textbf{36.80}/\textbf{0.866} \\
\cline{1-2}\cline{4-6}\cline{8-10}
\multirow{4}{*}{{Supervised}} & FBP+U-Net && 37.05/0.970 & \blue{30.26/0.723} & 32.13/0.820 && 32.13/0.936 & \blue{20.09/0.347} & 27.22/0.694 \\
&LGD && 40.73/0.985 & \blue{31.37/0.909} & 33.42/0.862 && 33.72/0.952 & \blue{22.86/0.687} & 28.49/0.507 \\
&LPD && \textbf{44.27}/\textbf{0.994} & \blue{28.09/0.918} & 33.21/0.866 && \textbf{36.19}/\textbf{0.970} & \blue{24.86/0.886} & 34.60/0.838 \\
&BDGD && \textbf{43.60}/\textbf{0.994} & \blue{30.72/0.974} & 35.91/0.877 && \textbf{35.36}/\textbf{0.971} & \blue{19.44/0.406} & 34.16/0.824\\
\cline{1-2}\cline{4-6}\cline{8-10}
\multicolumn{2}{c}{{BDGD+UKT}} && -- & \blue{\textbf{40.72}/\textbf{0.997}} &\textbf{38.40}/\textbf{0.899} && -- & \blue{\textbf{30.07}/\textbf{0.966}} & \textbf{35.67}/\textbf{0.855}\\
\cline{1-2}\cline{4-6}\cline{8-10}
\end{tabular}
}
}
\end{table*}

\begin{table*}[ht]
\begin{minipage}{.49\linewidth}
\centering
\caption{``Upper-bounds" obtained via supervised fine-tuning on LoDoFanB.}\label{tab:tab-supervised}
{\renewcommand{\arraystretch}{1.25}\small \setlength{\tabcolsep}{2pt}
\begin{tabular}{ccccc}
\cline{1-1}\cline{3-3}\cline{5-5}
\multicolumn{1}{c}{\boldtable{Methods}} && \boldtable{Low-Dose \ac{CT}} && \boldtable{Sparse-View \ac{CT}}\\
\cline{1-1}\cline{3-3}\cline{5-5}
\ac{FBP}+U-Net && 36.05/0.879 && 34.47/0.828\\
\ac{LGD} && 38.33/0.894 && 36.00/0.855  \\
\ac{LPD} && \textbf{39.85}/\textbf{0.914} && \textbf{37.59}/\textbf{0.876}\\
\ac{BDGD}+SKT && \textbf{40.14}/\textbf{0.909} && \textbf{37.71}/\textbf{0.877}\\
\cline{1-1}\cline{3-3}\cline{5-5}
\end{tabular}
}
\end{minipage}%
\quad
\begin{minipage}{.49\linewidth}
\centering
\caption{Comparison between \ac{BDGD}+\ac{UKT} and UL.}\label{tab:tab-UKT}
{\renewcommand{\arraystretch}{1.25}\small \setlength{\tabcolsep}{2pt}
\begin{tabular}{ccccc}
\cline{1-1}\cline{3-3}\cline{5-5}
\multicolumn{1}{c}{\boldtable{Methods}} && \boldtable{Low-Dose \ac{CT}} && \boldtable{Sparse-View \ac{CT}}\\
\cline{1-1}\cline{3-3}\cline{5-5}
\ac{BDGD}+\ac{UKT} && \textbf{38.33}/\textbf{0.895} && 35.67/\textbf{0.853}\\
\blue{\ac{UKT} w/o TV} && \blue{27.65/0.549} && \blue{22.86}/\blue{0.354}\\
U-\ac{BDGD} && 36.64/0.870 && \textbf{35.68}/0.852\\
\cline{1-1}\cline{3-3}\cline{5-5}
\end{tabular}
}
\end{minipage}
\end{table*}

Table \ref{tab:tab1} reports also the approximate runtime for all the methods under consideration.
All learned methods (\ie \ac{LGD}, \ac{LPD}, \ac{BDGD}) require multiple calls of the forward operator $A$, and thus they are slower at test time than the methods that do not (\eg \ac{FBP}+U-Net, which only post-processes the FBP reconstruction).
In addition, \ac{BDGD} and \ac{BDGD}+\ac{UKT} use 10 Monte Carlo samples to obtain a single reconstruction, leading to a slightly longer reconstruction time of approximately 7s per image.
However, all learned methods are found to be significantly faster than the TV reconstruction.
Meanwhile, \ac{DIP}+TV is much slower than TV taking approximately 20 minutes to reconstruct a single instance of the LoDoFanB dataset. \blue{The runtimes for the FoamFanB dataset were almost identical and are thus not included.}

Example reconstructed images are shown in {Figs.~\ref{fig:sparse_view} and  \ref{fig:low_dose}, for the sparse-view FoamFanB and the low-dose LoDoFanB} \ac{CT} settings, respectively. We observe that \ac{BDGD}+\ac{UKT} significantly reduces background noise in the reconstructions while faithfully capturing finer details, particularly in the low-dose setting.
Overall, \ac{DIP}+TV and \ac{BDGD}+\ac{UKT} produce reconstructions with similar properties. However, \ac{DIP}+TV, \ac{LPD} and \ac{BDGD}+\ac{UKT} tend to suffer from slight over-smoothing.
Meanwhile, TV reconstruction suffers from patchy artefacts, which is a well-known drawback of the TV penalty \cite{BrediesHoller:2020}, and also retains more background noise.

The sparse-view setting in Fig.~\ref{fig:sparse_view} is numerically more challenging and the reconstructions are susceptible to streak artefacts, which are especially pronounced in the \ac{FBP} but are still discernible in reconstructions obtained by other methods. Nonetheless, best performing methods (\ac{DIP}+TV and \ac{BDGD}+\ac{UKT}) can achieve an excellent compromise between smoothing and the removal of streak artefacts. 
Interestingly, in Fig \ref{fig:low_dose},
the learned methods, including \ac{BDGD}+\ac{UKT}, suffer from some undesirable over-smoothing inside the lung cavity.

\blue{\ac{BDGD}+\ac{UKT} is good at recovering fine structures that are present in the FoamFanB data, which are poorly reconstructed by \ac{BDGD}.
For example, in the last row of Fig.~\ref{fig:sparse_view}, the smaller circles are smoothed out and thus not discernible in \ac{BDGD} reconstructions, but they are well reconstructed with \ac{BDGD}+\ac{UKT}.
Similarly, Fig.~\ref{fig:low_dose} shows that \ac{BDGD}+\ac{UKT} better captures fine details in the human torso; for example the zoomed-in region shows an improvement over the overly-smoothed reconstruction produced by \ac{BDGD}. These observations clearly indicate that the unsupervised fine-tuning is highly beneficial in improving the quality of the reconstructed image.
}

We further evaluate the learned methods by first pretraining them on the Ellipses dataset, and then fine-tuning them on one half of the LoDoFanB dataset but with ground truth data included.
The remaining half of the LoBaFanB dataset is used for testing.
We thus operate under the assumption that we have access to only one half of the ground truth images from the LoDoFanB dataset.
This is intended to benchmark the reconstructive properties of the unsupervised fine-tuning against a more popular supervised adaptation, and may serve as the \blue{``upper-bound''} on the reconstructive performance of the proposed method. The quantitative results of this controlled setting are presented in Tables \ref{tab:tab-supervised} and \ref{tab:tab-UKT}.
The notation SKT stands for the supervised knowledge-transfer: the fine-tuning is conducted via \eqref{eqn:supervised_loss} on one half of the LoDoFanB dataset including ground truth data.
Unsupervised (U)-BDGD refers to \ac{BDGD} trained via \eqref{eqn:unsupervised_loss} by completely omitting the pretraining in the first phase.
It is observed that U-\ac{BDGD} shows subpar reconstructive properties only for the low-dose CT setting, but surprisingly, it matches the performance obtained by \ac{BDGD}+\ac{UKT} for the sparse-view \ac{CT} setting.
However, we observe that pretraining helps to considerably speed up the convergence of \ac{BDGD}+\ac{UKT}. It takes only a few epochs to converge, whilst U-\ac{BDGD} leads to a more unstable and lengthy learning (up to 100 epochs).
This behaviour is also observed with the fine-tuning of other learned benchmark methods.
This indicates the need of an adaptation phase, in the presence of distributional shift, and the beneficial effect of pretraining. Moreover, Table \ref{tab:tab-supervised} shows that using supervised data pairs from the target domain to adapt the network to the target task can significantly improve the reconstructive properties of all the learned methods.
Nonetheless, the degree of improvement depends strongly on both the used method and the problem setting.
The proposed \ac{BDGD}+\ac{UKT} approach dramatically improves the performance and mitigates the performance drop due to the distributional shift. \blue{Table \ref{tab:tab-UKT} also shows the influence of the TV in the fine-tuning stage. Setting $\gamma=0$ leads to overfitting to the noise after 10 epochs (i.e., approx. 2000 gradient updates), and even with careful early stopping the performance is still subpar when compared with the approach employing TV regularisation. Therefore, the TV term plays an important role in the proposed framework.}

{
\newcommand{\showpic}[2]{%
\begin{tikzpicture}[spy using outlines={rectangle, magnification=2.25,
size=2.75cm, connect spies, red, thick}]%
\draw (0, -4.95cm) node [anchor=south, yshift=-5pt] {\phantom{f}#1\phantom{g}};%
\draw (0,0) node [anchor=north] {\includegraphics[width=4cm, height=4cm,  angle=-90]{../figures/#2}};%
\spy on (0.25cm, -2cm) in node at (-0.75cm, -0.5cm);
\end{tikzpicture}\hspace*{-2mm}%
}

\begin{figure*}[h!]
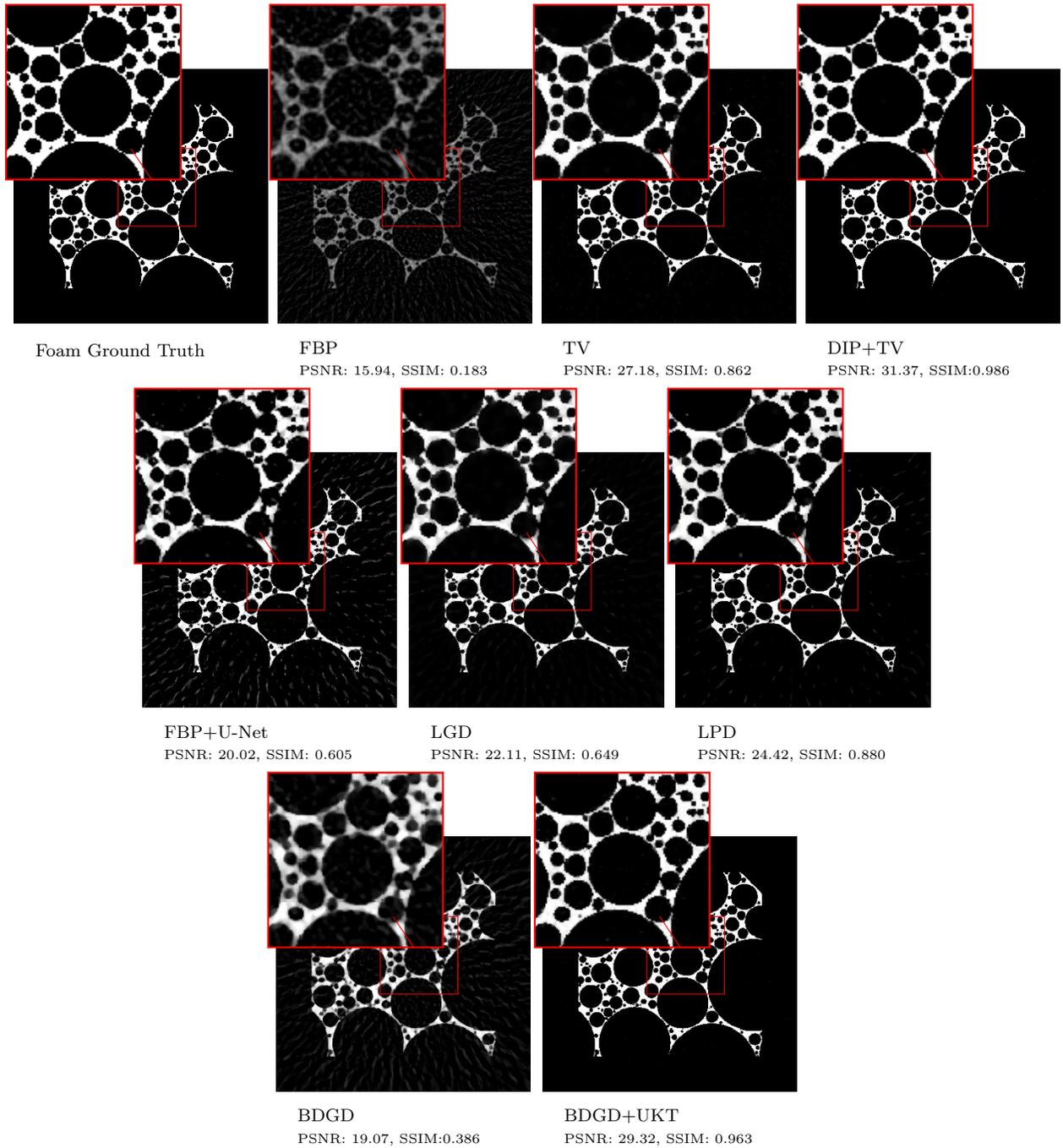

\centering
\quad\showpic{\parbox{3.25cm}{\scriptsize Foam Ground Truth\\}}{foam_slice0/sparse_view/foam_data_GT_0.png}
\showpic{\parbox{3.25cm}{\scriptsize FBP\\{\tiny PSNR: 15.94, SSIM: 0.183
}}}{foam_slice0/sparse_view/foam_data_FBP_0.png}
\showpic{\parbox{3.25cm}{\scriptsize TV\\{\tiny PSNR: 27.18, SSIM: 0.862
}}}{foam_slice0/sparse_view/foam_data_TV_0.png}
\showpic{\parbox{3.25cm}{\scriptsize DIP+TV\\{\tiny PSNR: 31.37, SSIM:0.986}}}{foam_slice0/sparse_view/foam_data_DIP+TV_0.png}
\quad
\\
\showpic{\parbox{3.25cm}{\scriptsize FBP+U-Net\\{\tiny PSNR: 20.02, SSIM: 0.605
}}}{foam_slice0/sparse_view/foam_data_UNET_0.png}
\showpic{\parbox{3.25cm}{\scriptsize LGD\\{\tiny PSNR: 22.11, SSIM: 0.649
}}}{foam_slice0/sparse_view/foam_data_DGD_0.png}
\showpic{\parbox{3.25cm}{\scriptsize LPD\\{\tiny PSNR: 24.42, SSIM: 0.880}}}{foam_slice0/sparse_view/foam_data_LPD_0.png}
\quad
\\
\showpic{\parbox{3.25cm}{\scriptsize BDGD\\{\tiny PSNR: 19.07, SSIM:0.386
}}}{foam_slice0/sparse_view/foam_data_BDGD_0.png}
\showpic{\parbox{3.25cm}{\scriptsize BDGD+UKT\\{\tiny PSNR: 29.32, SSIM: 0.963}}}{foam_slice0/sparse_view/foam_data_UKT_0.png}
\quad
\\
\vspace{.5em}
\caption{{Sparse-view reconstruction of the FoamFanB dataset along with a zoomed-in region indicated by a small square.}}
\label{fig:sparse_view}
\end{figure*}
}

{
\newcommand{\showpic}[2]{%
\begin{tikzpicture}[spy using outlines={rectangle, magnification=2.25,
size=2.75cm, connect spies, red, thick}]%
\draw (0, -4.95cm) node [anchor=south, yshift=-5pt] {\phantom{f}#1\phantom{g}};%
\draw (0,0) node [anchor=north] {\includegraphics[width=4cm, height=4cm,  angle=-90]{../figures/#2}};%
\spy on (0.25cm, -2cm) in node at (-0.75cm, -0.5cm);
\end{tikzpicture}\hspace*{-2mm}%
}

\begin{figure}[th!]
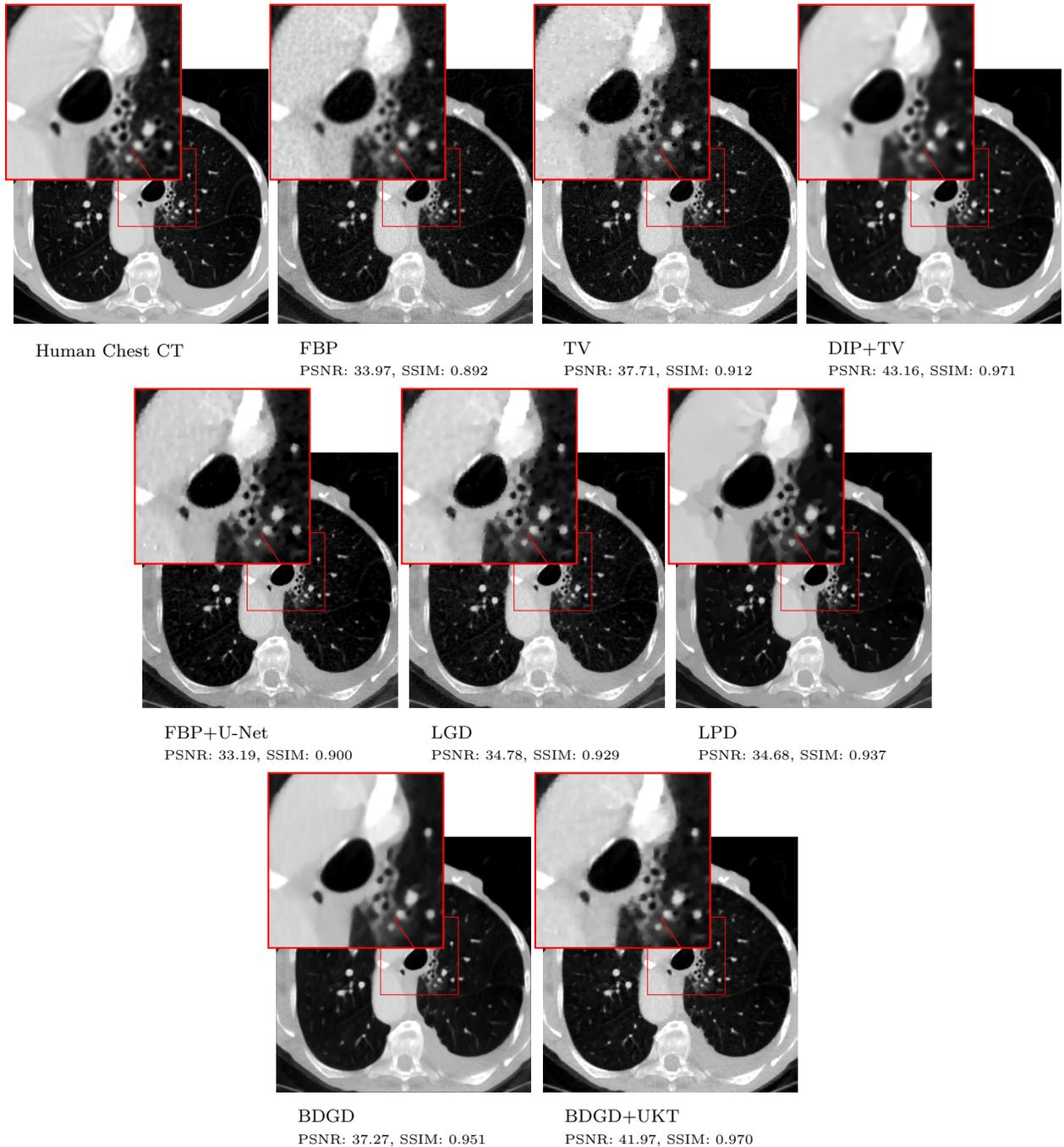

\centering
\quad
\showpic{\parbox{3.25cm}{\scriptsize Human Chest CT\\}}{lodofanb_slice7/low_dose/ground_truth}
\showpic{\parbox{3.25cm}{\scriptsize FBP\\{\tiny PSNR: 33.97, SSIM: 0.892}}}{lodofanb_slice7/low_dose/reco_fbp}
\showpic{\parbox{3.25cm}{\scriptsize TV\\{\tiny PSNR: 37.71, SSIM: 0.912}}}{lodofanb_slice7/low_dose/reco_tv}
\showpic{\parbox{3.25cm}{\scriptsize DIP+TV\\{\tiny PSNR: 43.16, SSIM: 0.971}}}{lodofanb_slice7/low_dose/reco_dip}
\quad
\\
\showpic{\parbox{3.25cm}{\scriptsize FBP+U-Net\\{\tiny PSNR: 33.19, SSIM: 0.900}}}{lodofanb_slice7/low_dose/reco_unet}
\showpic{\parbox{3.25cm}{\scriptsize LGD\\{\tiny PSNR: 34.78, SSIM: 0.929}}}{lodofanb_slice7/low_dose/reco_dgd}
\showpic{\parbox{3.25cm}{\scriptsize LPD\\{\tiny PSNR: 34.68, SSIM: 0.937}}}{lodofanb_slice7/low_dose/reco_lpd}
\quad
\\
\showpic{\parbox{3.25cm}{\scriptsize BDGD\\{\tiny PSNR: 37.27, SSIM: 0.951}}}{lodofanb_slice7/low_dose/reco_bdgd}
\showpic{\parbox{3.25cm}{\scriptsize \ac{BDGD}+\ac{UKT}\\{\tiny PSNR: 41.97, SSIM: 0.970}}}{lodofanb_slice7/low_dose/reco_ukt}
\quad
\\
\vspace{.5em}
\caption{Low-dose human chest \ac{CT} reconstruction within the LoDoFanB dataset along with a zoomed-in region indicated by a small square. The window is set to a  HU range of $\approx$ [-1000, 400].}
\label{fig:low_dose}
\end{figure}
}

{
\newcommand{\showpic}[2]{%
\begin{tikzpicture}[spy using outlines={rectangle, magnification=2.25,
size=2cm, connect spies, red, thick}]%
\draw (0, -4.25cm) node [anchor=south, yshift=-5pt] {\phantom{f}#1\phantom{g}};%
\draw (0,0) node [anchor=north] {\includegraphics[width=3.5cm, height=3.5cm, angle=-90]{../figures/#2}};%
\spy on (0.25cm, -2cm) in node at (-0.75cm, -0.5cm);
\end{tikzpicture}\hspace*{-2mm}%
}

\newcommand{\showpiclat}[3]{%
\begin{tikzpicture}[spy using outlines={rectangle, magnification=2.25,
size=2cm, connect spies, red, thick}]%
\draw (0, -4.25cm) node [anchor=south, yshift=-5pt] {\phantom{f}#1\phantom{g}};%
\draw (-1.85, -2.5cm) node [anchor=south, yshift=20pt, rotate=90] {\phantom{f}#2\phantom{g}};%
\draw (0,0) node [anchor=north] {\includegraphics[width=3.5cm, height=3.5cm,  angle=-90]{../figures/#3}};%
\spy on (0.25cm, -2cm) in node at (-0.75cm, -0.5cm);
\end{tikzpicture}\hspace*{-2mm}%
}

\newcommand{\colormap}[2]{%
\begin{tikzpicture}[]%
\draw (0cm, -1.0cm) node [anchor=south, yshift=-5pt] {\phantom{f}#1\phantom{g}};%
\draw (0,0) node [anchor=north] {\includegraphics[width=5cm, height=0.6cm]{../figures/#2}};%
\end{tikzpicture}\hspace*{-2mm}%
}

\begin{figure*}[!ht]
\centering
\vspace{-.75cm}
\begin{center}
    {\small Low-Dose CT}
\end{center}
\quad
\showpiclat{\parbox{3.25cm}{{\quad}\\}}{\scriptsize BDGD}{foam_slice0/low_dose/foam_data_error_no_ukt_0.png}
\showpic{\parbox{3.25cm}{{\quad}\\}}{foam_slice0/low_dose/foam_data_uq_no_ukt_0.png}
\showpic{\parbox{3.25cm}{{\quad}\\}}{foam_slice0/low_dose/foam_data_uqa_no_ukt_0.png}
\showpic{\parbox{3.25cm}{{\quad}\\}}{foam_slice0/low_dose/foam_data_uqe_no_ukt_0.png}
\quad
\\
\vspace{-.75cm}
\quad
\showpiclat{\parbox{3.25cm}{{\quad}\\}}{\scriptsize BDGD+UKT}{foam_slice0/low_dose/foam_data_error_ukt_0.png}
\showpic{\parbox{3.25cm}{{\quad}\\}}{foam_slice0/low_dose/foam_data_uq_ukt_0.png}
\showpic{\parbox{3.25cm}{{\quad}\\}}{foam_slice0/low_dose/foam_data_uqa_ukt_0.png}
\showpic{\parbox{3.25cm}{{\quad}\\}}{foam_slice0/low_dose/foam_data_uqe_ukt_0.png}
\quad
\\
\vspace{-.75cm}
\begin{center}
    {\small Sparse-View CT}
\end{center}
\quad
\showpiclat{\parbox{3.25cm}{{\quad}\\}}{\scriptsize BDGD}{foam_slice0/sparse_view/foam_data_error_no_ukt_0.png}
\showpic{\parbox{3.25cm}{{\quad}\\}}{foam_slice0/sparse_view/foam_data_uq_no_ukt_0.png}
\showpic{\parbox{3.25cm}{{\quad}\\}}{foam_slice0/sparse_view/foam_data_uqa_no_ukt_0.png}
\showpic{\parbox{3.25cm}{{\quad}\\}}{foam_slice0/sparse_view/foam_data_uqe_no_ukt_0.png}
\quad
\\
\vspace{-.75cm}
\quad
\showpiclat{\parbox{3.25cm}{\scriptsize Absolute Error}}{\scriptsize BDGD + UKT}{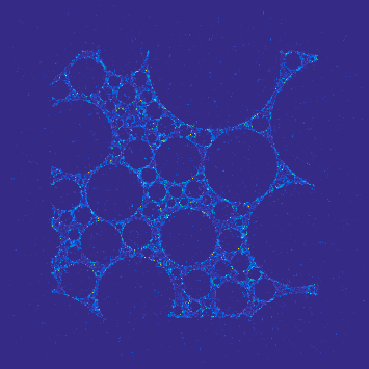}
\showpic{\parbox{3.25cm}{\scriptsize Overall Uncertainty}}{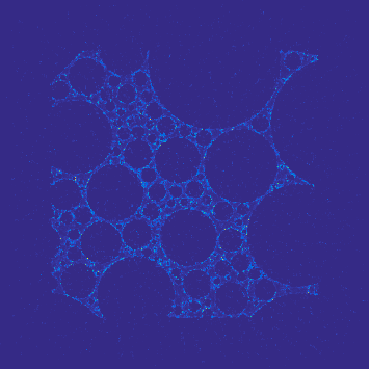}
\showpic{\parbox{3.25cm}{\scriptsize Aleatoric Component}}{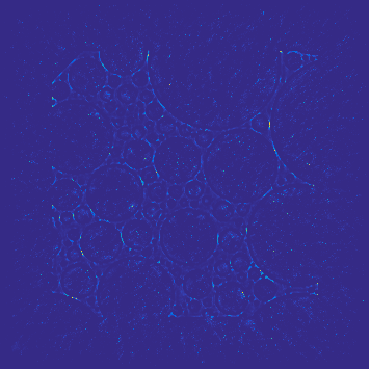}
\showpic{\parbox{3.25cm}{\scriptsize Epistemic Component}}{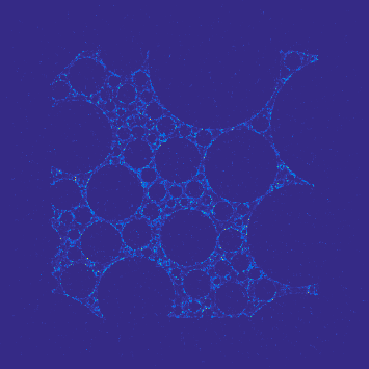}
\quad
\\
\quad
\colormap{\scriptsize Absolute Error/Uncertainty}{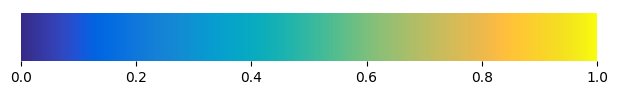}
\quad
\caption{{Qualitative uncertainty analysis on the FoamFanB dataset. The pixel-wise absolute reconstruction error, (max-min normalised across low-dose and sparse-view \ac{CT} settings) pixel-wise predictive uncertainty, and its decomposition into the aleatoric and epistemic constituent components for low-dose and sparse-view \ac{CT} obtained by \ac{BDGD} and \ac{BDGD}+\ac{UKT}.}}\label{fig:uq_figure_foam}
\end{figure*}
}

{
\newcommand{\showpic}[2]{%
\begin{tikzpicture}[spy using outlines={rectangle, magnification=2.25,
size=2cm, connect spies, red, thick}]%
\draw (0, -4.25cm) node [anchor=south, yshift=-5pt] {\phantom{f}#1\phantom{g}};%
\draw (0,0) node [anchor=north] {\includegraphics[width=3.5cm, height=3.5cm, angle=-90]{../figures/#2}};%
\spy on (0.25cm, -2cm) in node at (-0.75cm, -0.5cm);
\end{tikzpicture}\hspace*{-2mm}%
}

\newcommand{\showpiclat}[3]{%
\begin{tikzpicture}[spy using outlines={rectangle, magnification=2.25,
size=2cm, connect spies, red, thick}]%
\draw (0, -4.25cm) node [anchor=south, yshift=-5pt] {\phantom{f}#1\phantom{g}};%
\draw (-1.85, -2.5cm) node [anchor=south, yshift=20pt, rotate=90] {\phantom{f}#2\phantom{g}};%
\draw (0,0) node [anchor=north] {\includegraphics[width=3.5cm, height=3.5cm,  angle=-90]{../figures/#3}};%
\spy on (0.25cm, -2cm) in node at (-0.75cm, -0.5cm);
\end{tikzpicture}\hspace*{-2mm}%
}

\newcommand{\colormap}[2]{%
\begin{tikzpicture}[]%
\draw (0cm, -1.0cm) node [anchor=south, yshift=-5pt] {\phantom{f}#1\phantom{g}};%
\draw (0,0) node [anchor=north] {\includegraphics[width=5cm, height=0.6cm]{../figures/#2}};%
\end{tikzpicture}\hspace*{-2mm}%
}

\begin{figure*}[h!]
\centering
\vspace{-.75cm}
\begin{center}
    {\small Low-Dose CT}
\end{center}
\quad
\showpiclat{\parbox{3.25cm}{{\quad}\\}}{\scriptsize BDGD}{lodofanb_slice7/low_dose/abs_noukt}
\showpic{\parbox{3.25cm}{{\quad}\\}}{lodofanb_slice7/low_dose/600_std_wo_ukt.png}
\showpic{\parbox{3.25cm}{{\quad}\\}}{lodofanb_slice7/low_dose/600_ale_wo_ukt.png}
\showpic{\parbox{3.25cm}{{\quad}\\}}{lodofanb_slice7/low_dose/600_epi_wo_ukt.png}
\quad
\\
\vspace{-.75cm}
\quad
\showpiclat{\parbox{3.25cm}{{\quad}\\}}{\scriptsize BDGD+UKT}{lodofanb_slice7/low_dose/abs_ukt}
\showpic{\parbox{3.25cm}{{\quad}\\}}{lodofanb_slice7/low_dose/600_std_ukt.png}
\showpic{\parbox{3.25cm}{{\quad}\\}}{lodofanb_slice7/low_dose/600_ale_ukt.png}
\showpic{\parbox{3.25cm}{{\quad}\\}}{lodofanb_slice7/low_dose/600_epi_ukt.png}
\quad
\\
\vspace{-.75cm}
\begin{center}
    {\small Sparse-View CT}
\end{center}
\quad
\showpiclat{\parbox{3.25cm}{{\quad}\\}}{\scriptsize BDGD}{lodofanb_slice7/sparse_view/abs_noukt}
\showpic{\parbox{3.25cm}{{\quad}\\}}{lodofanb_slice7/sparse_view/100_std_wo_ukt.png}
\showpic{\parbox{3.25cm}{{\quad}\\}}{lodofanb_slice7/sparse_view/100_ale_wo_ukt.png}
\showpic{\parbox{3.25cm}{{\quad}\\}}{lodofanb_slice7/sparse_view/100_epi_wo_ukt.png}
\quad
\\
\vspace{-.75cm}
\quad
\showpiclat{\parbox{3.25cm}{\scriptsize Absolute Error}}{\scriptsize BDGD + UKT}{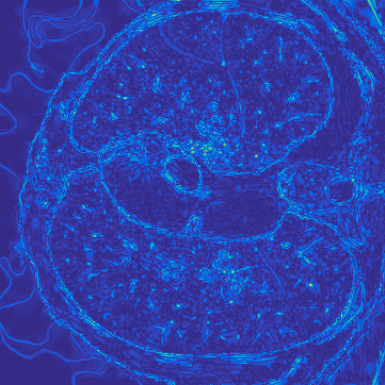}
\showpic{\parbox{3.25cm}{\scriptsize Overall Uncertainty}}{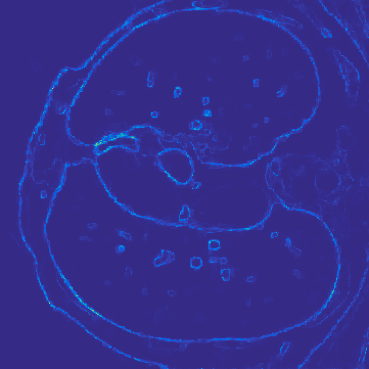}
\showpic{\parbox{3.25cm}{\scriptsize Aleatoric Component}}{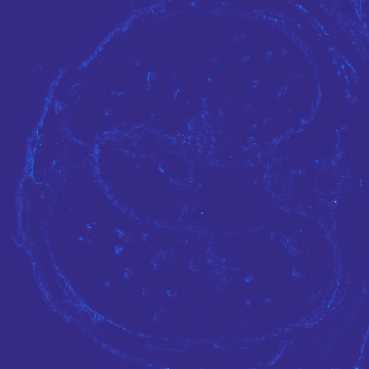}
\showpic{\parbox{3.25cm}{\scriptsize Epistemic Component}}{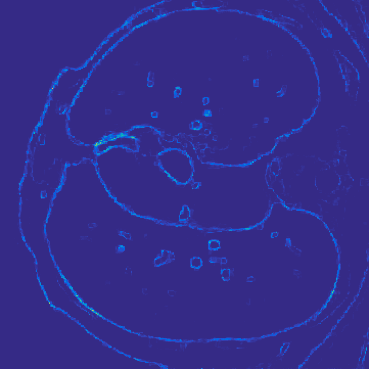}
\quad
\\
\quad
\colormap{\scriptsize Absolute Error}{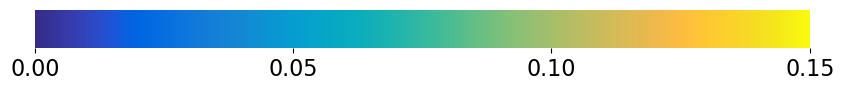}
\colormap{\scriptsize Uncertainty}{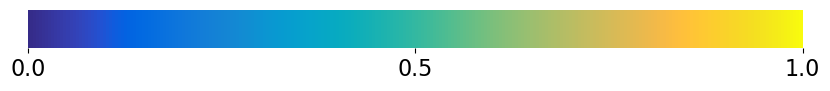}
\quad
\caption{Qualitative uncertainty analysis on the LoDoFanB dataset. The pixel-wise absolute reconstruction error, (max-min normalised across low-dose and sparse-view \ac{CT} settings) pixel-wise predictive uncertainty, and its decomposition into the aleatoric and epistemic constituent components for low-dose and sparse-view \ac{CT} obtained by \ac{BDGD} and \ac{BDGD}+\ac{UKT}.}\label{fig:uq_figure_lodofanb}
\end{figure*}
}

It is worth noting that \ac{BDGD}+\ac{UKT} also provides useful predictive uncertainty information on the reconstructions. In {Figs.~\ref{fig:uq_figure_foam} and \ref{fig:uq_figure_lodofanb}}, we present the uncertainty estimates along with pixel-wise errors for the FoamFanB and LoDoFanB \ac{CT} settings, respectively.
The overall predictive uncertainty largely concentrates around the edges: the reconstruction of sharp edges exhibits a higher degree of uncertainty.
This agrees well with the intuition that edges are more challenging to accurately resolve than smooth regions, and thus are more prone to reconstruction errors. Note that aleatoric and epistemic uncertainties have different sources, one is due to inherent data noise, and the other due to the model uncertainty, arising from the lack of a sufficient amount of training data. To ascertain the sources, we apply the decomposition \eqref{eqn:uncertainty_decomposition}. Interestingly, we observe that in both the low-dose and the sparse-view \ac{CT} settings, epistemic uncertainty appears to be dominating within the (overall) predictive uncertainty.
Nonetheless, the two types of uncertainty share a similar shape, and in either case, the overall shape closely resembles the pixel-wise error, indicating that the uncertainty estimate can potentially be used as an error indicator, concurring with existing empirical measurement data \cite{tanno2017bayesian}. It is also instructive to compare the uncertainty estimates obtained by \ac{BDGD} and \ac{BDGD}+\ac{UKT}. {Figs.~\ref{fig:uq_figure_foam} and~\ref{fig:uq_figure_lodofanb}} show that the estimates obtained by \ac{BDGD} result in larger magnitudes, with the aleatoric component overshadowing the epistemic one.
Visually, the unsupervised adaptation phase ameliorates the epistemic estimate: the pixel-wise predictive epistemic uncertainty obtained with \ac{BDGD}+\ac{UKT} is better at capturing the edges of the anatomical structures present in the reconstructed image.

\subsection{Discussion}\label{sec:discussion}

The experimental results in Tables \ref{tab:tab2} and \ref{tab:tab-UKT} have several implications for image reconstruction.
First, they show that while supervised iterative methods (\ac{FBP}+U-Net, \ac{LGD}, and \ac{LPD}) can deliver impressive results when trained and tested on imaging datasets that come from the same distribution, but fail when applied directly to data from a different distribution.
Specifically, on the Ellipses dataset they vastly outperform the traditional \ac{FBP} and TV, but on the LoDoFanB dataset the difference between learned methods and \ac{FBP} nearly vanishes (particularly in the low-dose setting), and the standard TV actually outperforms the supervised methods.
This behaviour might be due to a form of bias-variance trade-off, where training with a large dataset allows improving the performance in the supervised case, but which has a negative effect on the generalisation property. The performance degrades significantly when the distribution of the testing measurement data deviates from that of the training data.
This results in a loss of flexibility, and underwhelming performance, when reconstructing an image of a different type.
Thus, adjusting the training regiment, or further adapting the network parameters to data from a different distribution, can be beneficial for improving the reconstruction quality. The results in Table \ref{tab:tab-supervised} indicate that all learned methods, including \ac{BDGD}+\ac{UKT}, can benefit greatly from the supervised data from the target domain.

Overall, the results show that Bayesian neural networks with \ac{VI} can deliver strong performance that is competitive with deterministic reconstruction networks, when equipped with the strategy of \textit{being Bayesian only a little bit}.
This can be first observed on the Ellipses dataset. Table \ref{tab:tab1} shows that \ac{BDGD} performs on par with (or often better than) all the unsupervised and supervised methods under consideration, which is in agreement with previous experimental findings \cite{barbano2020quantifying}.
The results also show the potential of the Bayesian \ac{UKT} framework for medical image reconstruction in the more challenging setting where ground truth images are not available. Namely, adapting the model through the described framework allows achieving a significant performance boost on {both the FoamFanB} and LoDoFanB datasets.
Moreover, \ac{BDGD}+\ac{UKT} shows roughly the same performance as \ac{DIP}+TV, while being significantly faster in terms of runtime, cf. Table \ref{tab:tab1}. This observation is consistent with existing studies using pretraining in other contexts \cite{HeGirshick:2019,RaghuZhangBengio:2019}.
Indeed, all the learned methods are significantly faster than TV and \ac{DIP}+TV reconstructions. In addition, \ac{BDGD}+\ac{UKT} can deliver uncertainty estimates on the reconstructions, with their sources quantified into aleatoric and epistemic ones. It is observed that for the studied settings, the epistemic uncertainty dominates the aleatoric one, and both uncertainty estimates correlate well with the pixel-wise error of the reconstructions. Nonetheless, the calibration of these estimates remains to be validated, like nearly all DL-based uncertainty quantification techniques \cite{BarbanoArridgeJinTanno:2021}.

The extensive experimental results indicate that \ac{UKT} shows great promise in the unsupervised setting.
The results clearly show the need for adapting data-driven approaches to structural changes in the data, its distribution and size, and for incorporating the insights observed in the available supervised data to update the reconstruction model \cite{parisi2019continual, mundt2020wholistic}.
Though only conducted on labelling tasks, recent studies show that transfer learning through pretraining exhibits good results when the difference between data distributions is small \cite{yang2020transfer}.
Moreover, one needs to ensure that pretraining does not result in overfitting the data from the first task.
Both requirements seem to be satisfied in the studied setting.
Further investigation is needed to examine how does the performance of a reconstruction network change with respect to the size and type of data that the pretraining dataset consists of, as well as with respect to changes in the physical setting (\eg forward operators and noise statistics).

\section{Concluding Remarks}
\label{sec:conclusion}
The use of a full Bayesian  treatment for learned medical image reconstruction methods is still largely under development, due to the associated training challenges \cite{BarbanoArridgeJinTanno:2021}. The proposed \ac{BDGD}+\ac{UKT} is very promising in the following aspects: (i) it is easy to train due to the adoption of the strategy \textit{being Bayesian only a little bit}; (ii) the performance of the obtained point estimates is competitive with benchmark methods; (iii) it also delivers predictive uncertainty. In particular, the numerical results indicate that the predictive uncertainty can be visually used as a reliable error indicator.
In this work we have presented a novel two-phase learning framework, termed \ac{UKT}, for addressing the lack of a sufficiently large amount of paired training data in learned image reconstruction techniques. The framework consists of two learning phases, both within a Bayesian framework. It first pretrains a learned iterative reconstructor on (simulated) ordered pairs and then at test-time, it fine-tunes the model to realise sample-wise adaptation using only noisy clinically realistic measurements. Extensive experiments on low-dose and sparse-view \ac{CT} reconstructions show that the approach is indeed very promising. It can achieve competitive performance with several state-of-the-art supervised and unsupervised approaches both qualitatively and quantitatively.

\section*{Acknowledgements}
The authors are grateful to two anonymous referees for their constructive comments, and Johannes Leuschner (University of Bremen) for helpful discussions.

\appendix

\section{Alternative Loss for Unsupervised Knowledge-Transfer}

The derivation of the training loss \eqref{eqn:unsupervised_loss} is not very principled in the sense that the trace term $\mathrm{trace}(A\hat\Sigma A^\top)$ does not arise naturally from the likelihood function $p(y^\unsup|\theta_{\rm e})$ using Bayes' rule.
One can though derive alternative losses by slightly modifying the construction of the likelihood $p(y^\unsup|\theta_{\rm e})$.
Below we give one such construction based on hierarchical modelling. For any measurement datum $y^\unsup\in\data^\unsup$, corresponding to the unknown image $x^\unsup$, the likelihood $p(y^\unsup|x^\unsup)$ is set to
\[
p(y^\unsup|x^\unsup)=\CN(y^\unsup; A x^\unsup, \sigma^2 I),
\]
where $\sigma^2$ dictates the strength of the measurement noise. The likelihood $p(y^\unsup|x^\unsup)$ can be obtained under the standard assumption that the noise corruption to the exact data $Ax^\unsup$ (for the unknown image $x^\unsup$) follows a Gaussian distribution with zero mean and variance $\sigma^2I$. Meanwhile, under the heteroscedastic noise modelling, the unknown image $x^\unsup$ (which in turn depends on the network parameter $\theta$) is assumed to follow a Gaussian distribution
\begin{equation*}
p(x^\unsup|\theta_{\rm e})=\CN\big(x^\unsup; \net_{\theta}^\mu(x^\unsup_{0}), \hat{\Sigma}\big),
\end{equation*}
with the mean $\net_{\theta}^\mu(x^\unsup_{0})$ and covariance $\hat\Sigma$ being the two outputs of the neural network $\net_\theta$. Consequently, assuming that the data noise and the unknown image $x^\unsup$ are independent, combining the last two identities using Bayes' rule leads to
\begin{equation*}
   p(y^\unsup|\theta_{\rm e})=\CN(y^\unsup; A \net_\theta^\mu(x_0^\unsup), A\hat\Sigma A^\top+\sigma^2 I).
\end{equation*}
Upon letting $\bar y^\unsup = A \net_{\theta}^\mu(x^\unsup_{0})$, we then have
\begin{equation*}
    \log p(y^\unsup|\theta_{\rm e}) = -\dfrac{1}{2}(\bar y^\unsup - y^\unsup)^\top (A\hat\Sigma A^\top+\sigma^2 I)^{-1}(\bar y^\unsup - y^\unsup)  -\dfrac{1}{2}\log(\det(A\hat\Sigma A^\top+\sigma^2 I))-\dfrac{m}{2}\log(2\pi).
\end{equation*}
In addition to enforcing data fidelity, we may also include the total variation penalty into the loss (to stabilise the training process). Finally, after expanding, relabelling, and ignoring the constant terms (in $\theta$), we obtain the following alternative loss at the second phase
\begin{align}
\tilde{\CL}^\unsup(\theta_{\rm d},q_\psi(\theta_{\rm e}))  &=\dfrac{1}{N^\unsup}\sum_{n=1}^{N^\unsup}\BE_{q_{\psi}(\theta_{\rm e})}\Big[\dfrac{1}{2}(A\net_\theta^\mu(x_{n,0}^\unsup)- y^\unsup)^\top (A\hat\Sigma_nA^\top+\sigma^2 I)^{-1}(A\net_\theta^\mu(x_{n,0}^\unsup) - y^\unsup)\nonumber\\
&\quad+\dfrac{1}{2}\log(\det(A\hat\Sigma_n A^\top+\sigma^2 I)) + \gamma{{\rm TV}(\net_{\theta}^{\mu}(x^\unsup_{n,0}))} \Big] +
\beta \KL \left [ q_{\psi}(\theta_{\rm e}) \| q^\sup_{\psi^\ast}(\theta_{\rm e}) \right ].
\label{eqn:unsupervised_loss-alt}
\end{align}
Note that the term $\KL [ q_{\psi}(\theta_{\rm e}) \| q^\sup_{\psi^\ast}(\theta_{\rm e})]$ has a closed-form expression.

The loss in \eqref{eqn:unsupervised_loss-alt} differs from that in \eqref{eqn:unsupervised_loss} only in the construction of the likelihood $p(y^\unsup|\theta_{\rm e})$. However, the former is computationally less convenient, due to the presence of the factor $(A\hat\Sigma A^\top+\sigma^2 I)^{-1}$ in the data consistency term, as well as the log-determinant $\log(\det(A\hat\Sigma A^\top+\sigma^2 I))$. Indeed, in view of the following well-known matrix directional derivative formulas
\begin{align*}
   \frac{{\rm d}\log(\det(X))}{{\rm d} X} [H] = \mathrm{trace}(X^{-1} H)\quad \mbox{and}\quad  \frac{{\rm d}X^{-1}}{{\rm d} X}[H] = - X^{-1}H X^{-1},
\end{align*}
for any symmetric positive definite $X$, and admissible direction $H$,
the gradient evaluation requires solving multiple linear systems, with the matrices given only implicitly. This can be computationally demanding for large-scale image restoration tasks such as CT reconstruction. In practice, the derivative of the log-determinant can be efficiently approximated using randomised trace estimators (\eg the Hutchinson's estimator \cite{bujanovic2021norm}, which again involves multiple linear solves).

The next result shows that the loss in \eqref{eqn:unsupervised_loss} is actually a computationally more tractable  approximation to the genuine Bayesian loss in \eqref{eqn:unsupervised_loss-alt}, under the condition $A\hat{\Sigma}A^\top \ll \sigma^2I$ (\ie the matrix $\sigma^{-2}A\hat\Sigma A^\top$ has a small operator norm). This result provides a more principled Bayesian interpretation of the loss \eqref{eqn:unsupervised_loss}.
\begin{proposition}
Under the condition $A\hat{\Sigma}A^\top \ll \sigma^2I$, the loss in \eqref{eqn:unsupervised_loss} is an approximation to the Bayesian loss in \eqref{eqn:unsupervised_loss-alt}.
\end{proposition}
\begin{proof}
Let $r=A\net_\theta^\mu(x_0^\unsup)-y^\unsup$ be the residual. It follows directly from the preceding matrix derivative formulas that
\begin{align*}
    r^\top (A\hat\Sigma A^\top+\sigma^2 I)^{-1}r & \approx r^\top \sigma^{-2} Ir - r^\top\sigma^{-4}A\hat\Sigma A^\top r,\\
    \log(\det(A\hat\Sigma A^\top+\sigma^2I))&=\log( \det(\sigma^2I))+\log(\det(I+\sigma^{-2}A\hat\Sigma A^\top))  \\
    & \approx m\log \sigma^{2} + \mathrm{trace}(\sigma^{-2}A\hat\Sigma A^\top).
\end{align*}
Note that the approximation is good under the given conditions. Then substituting these approximations into \eqref{eqn:unsupervised_loss-alt}, ignoring the constant term and relabelling, we obtain the loss in \eqref{eqn:unsupervised_loss}. This shows the desired assertion.
\end{proof}

\bibliographystyle{abbrv}
\bibliography{reference}
\end{document}